# False discovery rate control under reduced precision computation for analysis of neuroimaging data

Hien D. Nguyen*, Yohan Yee,

Geoffrey J. McLachlan, and Jason P. Lerch

July 16, 2018

## Abstract

The mitigation of false positives is an important issue when conducting multiple hypothesis testing. The most popular paradigm for false positives mitigation in high-dimensional applications is via the control of the false discovery rate (FDR). Multiple testing data from neuroimaging experiments can be very large, and reduced precision storage of such data is often required. Reduced precision computation is often a problem in the analysis of legacy data and data arising from legacy pipelines. We present a method for FDR control that is applicable in cases where only $p$-values or test statistics (with common and known null distribution) are available, and when those $p$-values or test statistics are encoded in a reduced precision format. Our method is based on an empirical-Bayes paradigm

---

*HDN is at the Department of Mathematics and Statistics, La Trobe University, Bundoora 3086, Victoria Australia (Corresponding author; email: h.nguyen5@latrobe.edu.au). GJM is at the School of Mathematics and Physics, University of Queensland, St. Lucia 4072, Queensland Australia. YY and JPL are at the Mouse Imaging Centre, Hospital for Sick Children, M5T 3H7 Toronto, Ontario Canada.

where the probit transformation of the $p$-values (called the $z$-scores) are modeled as a two-component mixture of normal distributions. Due to the reduced precision of the $p$-values or test statistics, the usual approach for fitting mixture models may not be feasible. We instead use a binned-data technique, which can be proved to consistently estimate the $z$-score distribution parameters under mild correlation assumptions, as is often the case in neuroimaging data. A simulation study shows that our methodology is competitive when compared with popular alternatives, especially with data in the presence of misspecification. We demonstrate the applicability of our methodology in practice via a brain imaging study of mice.

# 1 Introduction

Modern experiments in numerous fields of science now output the results of thousands to millions of hypothesis tests simultaneously. Recent accounts of the theoretical aspects of the phenomenon of simultaneous statistical inference with applications in the life sciences can be found in Dudoit & van der Laan (2008) and Dickhaus (2014). Further treatments on the topic can be found in Efron (2010) and Efron & Hastie (2016, Ch. 15).

We assume that we are operating in a scenario whereupon we (only) observe $n \in \mathbb{N}$ $p$-values from $n$ simultaneous tests of the hypotheses $H_i$ ($i \in [n]$; $[n] = \{1, ..., n\}$), which may be either null or otherwise and may be related in some manner. Suppose that we are conducting well-specified standard significance tests at significance level $\alpha \in (0, 1)$. If all of the hypotheses are null, then we can directly compute the expected number of tests declared significant as $n\alpha$. Taking $n$ large (e.g. $n \geq 10^6$) and $\alpha$ at usual levels such as $\alpha \in (0.001, 0.1)$, the number of incorrectly declared hypotheses as not null can be greatly inflated. In the case where not all hypotheses are null, the outcome of using only standard significant tests can lead to absurd conclusions, such as identifying neuronal activation in the brain of a dead fish via functional magnetic resonance imaging (Bennett et al., 2009).

In recent years, the leading paradigm for the handling of large-scale simultaneous hypothesis testing scenarios is via the control of the false discovery rate (FDR) of an experiment. The control of FDR was first introduced by Benjamini & Hochberg (1995) and has been developed upon by numerous other authors. The FDR of an experiment can be defined as $\text{FDR} = \mathbb{E}(N_{01}/N_R) \mathbb{P}(N_R > 0)$, where $N_{01}$ and $N_R$ denote the number of false positives and the number of rejected hypotheses (hypotheses declared significantly alternative) from the experiment, respectively.

The FDR control method of Benjamini & Hochberg (1995) was first developed to only take an input of $n$ IID (identically and independently distributed) $p$-values. An extension towards the control of FDR in samples of correlated $p$-values was derived in Benjamini & Yekutieli (2001). Since these key publications, there have been numerous articles written on the topic of FDR control in various settings and under various conditions; see Benjamini (2010) and the comments therein for an account of the history and development of FDR control. A classic treatment regarding FDR control in neuroimaging can be found in Genovese et al. (2002).

In most FDR control methods, there is an explicit assumption that the marginal distribution of the $p$-values of an experiment is uniform over the unit interval, if the hypothesis under consideration is null. This assumption arises via the classical theory of $p$-values of well-specified tests (cf. Dickhaus, 2014, Sec. 2). However, in practice, there are numerous ways for which the distri-bution of $p$-values under the null can deviate from uniformity. In Efron (2010, Sec. 6.4), several causes of deviation from uniformity are suggested. Broadly, these are: failed mathematical assumptions (e.g. incorrect use of distribution for computing $p$-values), correlation between $p$-values, and unaccounted covariates or misspecification of null hypotheses. A treatment on the effects of misspecification of the null hypotheses due to unaccounted covariates can be found in Barreto & Howland (2006, Ch. 7 Appendix and Ch. 18).

There are some FDR methods that account for deviation from uniformity in the null distribution. These include the methods of Yekutieli & Benjamini (1999), Korn et al. (2004), Pollard & van der Laan (2004), van der Laan & Hubbard (2006), and Habiger & Pena (2011). Unfortunately, the listed methods all require access to the original data of the experiment in order to compute permutation-based test statistics and thus permutation-based $p$-values. As mentioned previously, access to the original experimental data lies outside of the scope of this article as we only assume knowledge of the $p$-values.

Fortunately, the empirical-Bayes (EB) paradigm of Efron (2010) provides a powerful framework under which the deviation of the null away from uniformity can be addressed with only access to the experimental $p$-values. For $i \in [n]$, let $P_i$ be a $p$-value and let $Z_i = \Phi^{-1}(1 - P_i)$ be the probit transformation of $P_i$. We refer to $Z_i$ as the $z$-scores. Here $\Phi$ is the cumulative distribution function of the standard normal distribution. Under the EB paradigm, we assume that some $\pi_0 \in [0, 1]$ proportion of the $n$ hypotheses are null and thus $\pi_1 = 1 - \pi_0$ are otherwise. Since an alternative (not null) hypothesis generates a $p$-value that is on average smaller than that of a null hypothesis, we can also assume that the $z$-scores of null hypotheses arise from some distribution with a mean $\mu_0 \in \mathbb{R}$, where $\mu_0 < \mu_1$ and $\mu_1 \in \mathbb{R}$ is the mean of the alternative $z$-scores. Since under uniformity of the $p$-values, the $z$-scores have a standard normal distribution, we can approximate the density of the null $z$-scores by $f_0(z) = \phi(z; \mu_0, \sigma_0^2)$, where $\sigma_0^2 > 0$ and $\phi(\cdot; \mu, \sigma^2)$ is the normal density function with mean $\mu$ and variance $\sigma^2$. Likewise, we can approximate the density of the alternative $z$-scores by $f_1(z) = \phi(z; \mu_1, \sigma_1^2)$, where $\sigma_1^2 > 0$ (cf. Efron, 2004). Thus, the marginal density of any $z$-score, whether null or otherwise, can be approximated by the two-component mixture model

$$f(z; \boldsymbol{\theta}) = \pi_0 f_0(z) + \pi_1 f_1(z), \qquad (1)$$

where $\boldsymbol{\theta}^\top = (\pi_0, \mu_0, \sigma_0^2, \mu_1, \sigma_1^2)$ is the model parameter vector and $(\cdot)^\top$ is the transpose operator; see McLachlan & Peel (2000) for details regarding finite mixture models.

The EB paradigm for FDR control was first introduced in Efron et al. (2001) and has been developed upon in articles such as Efron & Tibshirani (2002), Efron (2004), Efron (2007a), Efron (2007b), and Efron & Hastie (2016, Ch. 15); see also the works of McLachlan et al. (2006), Jin & Cai (2007), Sun & Cai (2007), Xie et al. (2011), and Nguyen et al. (2014). A relatively complete account of the EB paradigm appears in Efron (2010).

We largely follow the works of McLachlan et al. (2006) and Nguyen et al. (2014). Our novelty and development on the available literature is to present a methodology for addressing the problems that are induced by data compression and storage algorithms that are commonly used in neuroimaging. Neuroimaging data such as MRI and functional MRI volumes are usually stored via one of a number of common storage protocols.

These protocols include ANALYZE (Robb et al., 1989), DICOM Bidgood et al. (1997), MINC (Vincent et al., 2003), and NIFTI (Cox et al., 2004). A good summary of these protocols is presented in Larobina & Murino (2014). In the past, in the pursuit of reduced storage sizes, it is not uncommon for neuroimaging data volumes to be stored at the minimum precision specification of any of the aforementioned formats. For example, DICOM volumes can only store data as integers, at a precision level as low as 8-bits (i.e. $2^8 = 256$ unique values).

In this article, we deal exclusively with this use of the integer storage option of the volume storage protocols and their rounding and truncation effects on the $p$-values that are outputted from an experiment for which FDR control is required. The importance of our investigation arises due to the popularity of open-access data repositories such as those that are cited in Eickhoff et al. (2016). These data repositories offer a wealth of opportunities for new scientific discoveries. Unfortunately, the open-access data arise from such dated studies are often stored in legacy data formats. Such legacy data formats are generally low precision, even after conversion for analysis with modern pipelines, such as via the methods of Li et al. (2016). The analysis of legacy data from open-access repositories therefore methodological consideration.

It is known that quantization and compression of real numbers can often lead to inaccuracies in statistical computations. Discussions of some aspects regarding the effects of quantization on statistical computation are discussed in Vardeman & Lee (2005) and Moschitta et al. (2015). The effects of quantization are particularly ruinous when applying EB-based FDR control since the conversions between $p$-values with values zero or one to $z$-scores are not finite.

Due to the infinite nature of compression-effected $z$-scores, the $p$-values for which such $z$-scores are derived cannot be included in the estimation of the parameters of the $z$-score density $f$, via naive estimation techniques. As such, the only way that the parameter elements $\boldsymbol{\theta}$ can be estimated is by omitting the infinite $z$-scores, which may cause a bias in the parameter estimate. Furthermore, even when there are no infinite $z$-scores, there is a potential bias that is incurred due to the quantization of the $p$-values that are induced by the integer storage protocols. The effect of these compression schemes constitutes a failed mathematical assumption under the taxonomy of Efron (2010, Sec. 6.4).

In this article, we address the problem of reduced precision $p$-values that are effected by integer storage options, via the use of binned estimation. That is, the $z$-scores are binned into interval classes before the estimation of the parameter elements of $f$ occurs. The estimation of normal mixture models from binned data has been investigated in numerous publications such as: MacDonald & Green (1988), McLachlan & Jones (1988), Jones & McLachlan (1992), Cadez et al. (2002), Same (2009), Lee & Scott (2012), Wu & Hamdan (2013a), and Wu & Hamdan (2013b).

We firstly demonstrate the effect of reduced precision storage of $p$-values on the estimation of the null distribution $f_0$, when all hypotheses are null via a numerical study, and the effect of estimation of the $z$-score distribution $f$, when there is a mix of null and alternative hypotheses. Making use of the EM (expectation–maximization of Dempster et al., 1977) algorithm from the **mix** function in the **mixdist** package (MacDonald & Green, 1988; Du, 2002; MacDonald & Du, 2012) in the **R** programming language (R Core Team, 2016), we demonstrate that one can easily and quickly compute the parameter elements of the $z$-score density $f$ via maximum marginal likelihood (MML) estimation (cf. Varin, 2008). We further prove that the MML estimator for the parameter elements of $f$ can be estimated consistently, even when there are dependencies between the $p$-values. A second numerical study is conducted to demonstrate the performance of our method under compression of $p$-values, where a comparison between our method is made against the commonly used methods of Benjamini & Hochberg (1995) and Benjamini & Yekutieli (2001), as well as the $q$-value method of Storey (2002), which is closely related to EB-based FDR control (cf. Efron, 2010, Sec. 4.6). An example application to a mouse brain imaging dataset is then provided to show the usefulness of our approach in a real data scenario.

A parallel application of the EB paradigm for FDR control for neuroimaging data was considered by Bielczyk et al. (2018). In Bielczyk et al. (2018), the authors constructed a discriminating mixture on the space of effect sizes instead of $z$-scores computed from $p$-values, and utilized their methodology to control FDR in connectomics experiments. The methods that are presented in Bielczyk et al. (2018) do not address the problem of reduced precision computation that may arise from the analysis of legacy data from open-access repositories.

The article proceeds as follows. In Section 2, we present a numerical study on the effect integer storage on $p$-value$s$, and subsequently on the naive estimation of $f_0$ and $f$. In Section 3, MML estimation of the parameter vector of $f$ is discussed and the use of the estimated density $\hat{f}$ for EB-based FDR control is established. In Section 4, a numerical study of the performance of our method is presented. In Section 5, the methodology is applied to control the FDR of a mouse imaging data set. Conclusions are drawn in Section 6.

## 2 The Effects of Reduced Precision via Integer Storage

Table 1 of Larobina & Murino (2014) provides a summary of the possible data compression schemes that can be applied when storing data in the ANALYZE, DICOM, MINC, or NIFTI formats. The possible integer storage schemes available for ANALYZE are 8-bits unsigned, or 16 and 32-bits signed. For DICOM, the available schemes are 8, 16, and 32-bits signed or unsigned. For MINC, 8, 16, and 32-bits signed or unsigned, are available. Finally, NIFTI can store data as 8, 16, 32, or 64-bits signed or unsigned.

For reference, 8, 16, 32, and 64 binary bits unsigned can encode 256, 65536, 4294967296, and 1.84E+19 ($a\text{E}b = a \times 10^b$) unique values, respectively. These numbers are doubled when signed encodings are used. In this article, we only consider integer compression in 8-bits or 16-bits signed and unsigned formats. This is because 32-bits and 64-bits can be used to encode single and double-precision floating points, respectively, which largely mitigate against the reduced precision problems that we discuss in this article.

## 2.1 Integer Encoding of Data

We are largely concerned with large scale-hypothesis testing situations that arise from voxel-based experiments (cf. Ashburner & Friston, 2000). In such experiments, a hypothesis test is conducted at each voxel of an imaged volume. For statistical analyses, resulting volumes of $p$-values or test statistics (with common and known null distribution) are generated. It is these volumes that are then stored, possibly in a reduced precision format, for dissemination or for storage.

Suppose that a $\gamma$-bits unsigned integer encoding is used, where $\gamma \in \mathbb{N}$. Note that a $\gamma$-bits signed integer encoding is effectively equivalent to a $(\gamma + 1)$-bits unsigned, for all intents and purposes. When the hypothesis testing data are stored as a $p$-values volume, we suppose that the data are stored such that the smallest integer value encodes the number zero and the largest integer value encodes the number one. The remainder of the integers are used to encode the unit interval at equally-spaced points. The encoding process then rounds the original $p$-values towards the nearest of these equally-spaced points. We refer to this approach as a $\gamma$-bits $p$-type encoding. Under the storage protocols that we assess, $\gamma \in \{8, 9, 16, 17\}$ generate valid encodings.

Now suppose that a $\gamma$-bits signed integer encoding is used. Note that a $\gamma$-bits unsigned integer can be used to effectively simulate a $(\gamma - 1)$-bits signed encoding. When the testing data are stored as a volume of test statistics, we store the data symmetrically about the origin. Here, the smallest integer value encodes the number zero and the largest integer value encodes the largest number in the volume, in absolute value. The remainder of the integers are used to encode the interval between zero and the largest number, at equally-spaced points. The encoding process then rounds the absolute value of the original test statistics towards the nearest equally-spaced points, and also store the signs of the values. We call this approach a $\gamma$-bits $T$-type encoding. Under the storage protocols that we assess, $\gamma \in \{7, 8, 15, 16\}$ generate valid encodings.

*Remark* 1. In practice, any bounded positive interval of real numbers that is encoded as a $\gamma$-bits integer is scaled so that the smallest value equates to zero and the largest value of the bounded equates to $2^\gamma - 1$. After scaling, any number that is not an integer is rounded to the nearest integer. This scheme is equivalent to the $\gamma$-bits $p$-type encoding scheme, and is the typical compression output that is obtained as a result of integer encoding under any of the protocols from Larobina & Murino (2014). A similar comment can be made regarding the use of signed integer encodings.

## 2.2 The Effect on the Null Distribution

Let $H_i$ each be null for $i \in [n]$ and assume that each is tested using a well-specified test resulting in a $p$-value $P_i$ arising from a uniform distribution over the unit interval (cf. Dickhaus, 2014, Ch. 2). To simulate a well-specified test, we simulate test statistics $T_i$ from the standard normal distribution and compute the respective $p$-values under the null hypothesis that the test statistics arise from the standard normal distribution (from which they are generated). That is, we compute each $p$-value as $P_i = 1 - \Phi(T_i)$. Generating $p$-values in this way adheres to the theory from Dickhaus (2014, Ch. 2).

Table 1: Monte Carlo mean and standard errors (SE) from the average of 100 ML estimates $\hat{\mu}_0$ and $\hat{\sigma}_0^2$, based on $p$-type and $T$-type integer encodings of testing data.

|                    | $\hat{\mu}_0$ |           | $\hat{\sigma}_0^2$ |           |
|-------------------:|--------------:|----------:|-------------------:|----------:|
| Encoding           | Mean          | SE        | Mean               | SE        |
| None               | 3.45E-04      | 3.09E-03  | 1.00E+00           | 2.11E-03  |
| 8-bits $p$-type    | 4.90E-04      | 2.91E-03  | 9.63E-01           | 3.95E-03  |
| 9-bits $p$-type    | -4.29E-05     | 3.00E-03  | 9.79E-01           | 4.39E-03  |
| 16-bits $p$-type   | -2.37E-04     | 2.97E-03  | 9.99E-01           | 4.43E-03  |
| 17-bits $p$-type   | -1.31E-04     | 3.19E-03  | 1.00E+00           | 4.38E-03  |
| 7-bits $T$-type    | -1.02E-05     | 2.27E-03  | 1.00E+00           | 2.47E-03  |
| 8-bits $T$-type    | -3.22E-04     | 3.10E-03  | 1.00E+00           | 2.11E-03  |
| 15-bits $T$-type   | 1.34E-04      | 2.78E-03  | 1.00E+00           | 2.09E-03  |
| 16-bits $T$-type   | 1.32E-04      | 3.16E-03  | 1.00E00            | 2.47E-03  |

We simulate $n = 10^6$ test statistics and encode the simulated data using $\gamma$-bits $T$-type encodings for all valid values of $\gamma$. We also simulate $n = 10^6$ $p$-values and encode the simulated data using $\gamma$-bits $p$-type encodings for all valid values of $\gamma$. The respective $z$-scores from each encoding scenario are computed, and the parameter elements of $f_0(z) = \phi(z; \mu_0, \sigma_0^2)$ are then estimated via ML estimation. Here, we naively omit infinite $z$-scores. The process is repeated 100 times for each encoding rule. We also estimate the parameter elements of $f_0(z)$ for $n = 10^6$ $z$-scores that are obtained without encoding in order to provide a benchmark. All computations are conducted in **R**.

Table 1 contains the results from the numerical study that is set up above. In the table and elsewhere, we denote the estimate/estimator of any quantity $\theta$ as $\hat{\theta}$.

Theoretically, we would anticipate that there is no deviation away from a standard normal distribution when no encoding is introduced. This is exactly what we observe in the first row of Table 1, where neither the average of the mean nor variance estimates are outside of a 95% confidence interval (CI; i.e.

approximately Mean $\pm 2 \times$ SE). In fact, only two encoding schemes (8 and 9-bits $p$-type encodings) resulted in significant differences of any kind, from the anticipated estimated values.

In each case the estimated variance is reduced from the nominal value of $\sigma_0^2 = 1$. This reduction can be explained by the fact that the finite $z$-scores distributions that are obtained from the probit transformation of the encoded $p$-values are approximately standard normal distributions that are truncated to the interval $[-a_\gamma, a_\gamma]$, where $a_\gamma = \Phi^{-1}\left(1 - 1/\left[2^{\gamma+1} - 1\right]\right)$. Using the variance formula for a doubly truncated standard normal distributions (cf. Forbes et al., 2011, Sec. 33.4), we have the variance formula for the $z$-scores:

$$\mathrm{var}_\gamma = 1 - 2a_\gamma \phi\left(a_\gamma; 0, 1\right) / \left[\Phi\left(a_\gamma\right) - \Phi\left(-a_\gamma\right)\right].$$

Substituting 8 and 9 into $\gamma$, we obtain truncated variances of $\mathrm{var}_8 = 9.64\mathrm{E}\text{-}1$ and $\mathrm{var}_9 = 9.80\mathrm{E}\text{-}1$, respectively. These values are almost identical to those from Table 1.

*Remark* 2. We note that the extra $1/2$ factor in the calculation of $a_\gamma$ (i.e. $1/2^{\gamma+1}$ rather than $1/2^\gamma$) arises from the fact that half of the $p$-values in the interval between zero and the next smallest number gets rounded towards the zero, and similarly half of the $p$-values in the interval between one and the next largest number gets rounded towards one. Thus, we lose approximately $1/\left[2^\gamma - 1\right]$ observations from the extreme values of the $p$-values distribution that probit transform to infinite values. Here, the $-1$ term accounts for a fencepost error.

## 2.3 The Effect on the $z$-score Distribution

Now suppose that the hypotheses $H_i$ are generated from two populations, a null one with probability $\pi_0 = 0.8$, and an alternative one with probability $\pi_1 = 0.2$. Under the null hypothesis, we generate test statistics $T_i$ from a standard normal distribution, and under the alternative, we generate test statistics from a normal distribution with mean $\mu_1 = 2$ and variance $\sigma_1^2 = 1$, instead. The $p$-values $P_i = 1 - \Phi(T_i)$, for testing the null that the test statistics are standard normal, are also computed. Again, we let $n = 10^6$.

Encoding of the $p$-values or test statistics is again conducted under one of the protocols that are described in Section 2.1. We then compute $z$-scores and discard any infinite values. The parameter vector $\boldsymbol{\theta}$ is then estimated via ML estimation. The process is again repeated 100 times for each encoding type. ML estimation is conducted via the usual EM algorithm for finite mixtures of normal distributions via the **normalmixEM2comp** function from the package **mixtools** (Benaglia et al., 2009). The result of this numerical study is reported in Table 2.

There were no differences between the results of any $T$-type encoding when compared to the no encoding results. All of the $T$-type encoding and no encoding results yield estimator confidence intervals that are insignificantly different from the generative model that values that were declared, earlier. Therefore, we conclude that there is insufficient evidence to conclude that $T$-type encoding of testing data leads to harmful effect when used to estimate the EB-based FDR model that was considered in this study.

Unfortunately, the same conclusion could not be made for the $p$-type encodings. The estimated parameter elements were uniformly significantly different from the generative values for the model. As $\gamma$ increases, we observe that that estimated values appear to approach the nominal parameter values. However, this approach appears to be slow and still leads to significantly incorrect estimates, even for the largest considered $\gamma$.

We can again provide a reason for the incorrect results that are obtained from the $p$-type inference. Let $a_\gamma = \Phi^{-1}\left(1 - 1/\left[2^{\gamma+1} - 1\right]\right)$, as in Section 2.2.

Table 2: Mean and standard errors (SE) for the average of 100 ML estimates of $\boldsymbol{\theta}$, based on $p$-type and $T$-type integer encodings of testing data.

| Encoding | $\hat{\pi}_0$ Mean | SE | $\hat{\mu}_0$ Mean | SE | $\hat{\sigma}_0^2$ Mean | SE | $\hat{\mu}_1$ Mean | SE | $\hat{\sigma}_1^2$ Mean | SE |
|---|---|---|---|---|---|---|---|---|---|---|
| None | 7.88E-01 | 1.65E-02 | -1.52E-02 | 2.12E-02 | 9.90E-01 | 1.61E-02 | 1.95E+00 | 7.71E-02 | 1.04E+00 | 5.63E-02 |
| 8-bits $p$-type | 9.24E-01 | 4.48E-03 | 1.21E-01 | 9.17E-03 | 1.05E+00 | 1.02E-02 | 2.23E+00 | 2.90E-02 | 1.16E-01 | 1.11E-02 |
| 9-bits $p$-type | 9.21E-01 | 4.41E-03 | 1.38E-01 | 7.98E-03 | 1.09E+00 | 1.05E-02 | 2.33E+00 | 2.69E-02 | 1.59E-01 | 1.10E-02 |
| 16-bits $p$-type | 8.60E-01 | 6.42E-03 | 7.99E-02 | 1.10E-02 | 1.06E+00 | 1.29E-02 | 2.31E+00 | 3.67E-02 | 6.67E-01 | 2.27E-02 |
| 17-bits $p$-type | 8.49E-01 | 7.07E-03 | 6.33E-02 | 1.22E-02 | 1.05E+00 | 1.28E-02 | 2.25E+00 | 4.05E-02 | 7.35E-01 | 2.74E-02 |
| 7-bits $T$-type | 7.87E-01 | 1.62E-02 | -1.59E-02 | 2.01E-02 | 9.88E-01 | 1.53E-02 | 1.94E+00 | 7.67E-02 | 1.04E+00 | 5.47E-02 |
| 8-bits $T$-type | 7.89E-01 | 1.75E-02 | -1.31E-02 | 2.20E-02 | 9.91E-01 | 1.76E-02 | 1.95E+00 | 8.18E-02 | 1.03E+00 | 5.81E-02 |
| 15-bits $T$-type | 7.88E-01 | 1.67E-02 | -1.58E-02 | 2.17E-02 | 9.89E-01 | 1.69E-02 | 1.94E+00 | 7.73E-02 | 1.04E+00 | 5.59E-02 |
| 16-bits $T$-type | 7.88E-01 | 1.67E-02 | -1.44E-02 | 2.08E-02 | 9.90E-01 | 1.67E-02 | 1.94E+00 | 7.76E-02 | 1.04E+00 | 5.69E-02 |

The ML estimator is estimated using approximately

$$n \left( \pi_0 \left[ \Phi \left( a_\gamma \right) - \Phi \left( -a_\gamma \right) \right] + \pi_1 \left[ \Phi \left( a_\gamma - 2 \right) - \Phi \left( -a_\gamma - 2 \right) \right] \right) < n$$

observations from the distribution that is characterized by the density $f(z; \boldsymbol{\theta})$, that is truncated on the interval $[-a_\gamma, a_\gamma]$. Note that no member of the family of densities of form $f(z; \boldsymbol{\theta})$ can perfectly match a truncated version of the density. For example, the two families of densities have different supports. Thus, the ML estimation procedure results in an estimated set of parameter values that yields a member of the untruncated density that best approximates the truncated density, in Kullback-Leibler divergence (cf. White, 1982). This approximation process explains the difference between the estimated parameter values and the generative parameter values. The smaller sample size explains the larger standard errors that are observed, uniformly over the estimates of the parameter elements.

## 3 Binned Estimation of the $z$-score Distribution

Let $-\infty = b_0 < b_1 < b_2 < ... < b_{m-1}$ for some $m \in \mathbb{N} \setminus \{1\}$. We define $m$ bins $B_j$, for $j \in [m]$, where $B_j = (b_{j-1}, b_j]$ for $j \in [m-1]$ and $B_m = (b_{m-1}, \infty)$. Suppose that we observe $n$ $p$-values $P_i$ that are converted to $z$-scores $Z_i$ that may be potentially infinite. Further, define $\mathbb{I}(A)$ as the indicator variable that takes value 1 if proposition $A$ is true and 0 otherwise, and define a new random variable $\boldsymbol{X}_i^\top = (X_{i1}, ..., X_{im})$, where $X_{ij} = \mathbb{I}(Z_i \in B_j)$, for each $i$ and $j \in [m]$.

Suppose that the $n$ $p$-values generate $z$-scores that are potentially correlated and marginally arise from a mixture model of form (1), with $\boldsymbol{\theta} = \boldsymbol{\theta}^0$, for some valid $\boldsymbol{\theta}^0$. Using the bins and realizations $\boldsymbol{x}_i^\top = (x_{i1}, ..., x_{im})$ of each $\boldsymbol{X}_i$ ($i \in [n]$), we can write the marginal likelihood and log-marginal likelihood functions under the mixture model approximation for the $z$-scores as

$$L(\boldsymbol{\theta}) = \prod_{i=1}^{n} \prod_{j=1}^{m} \left[ \int_{B_j} f(z; \boldsymbol{\theta}) \, \mathrm{d}z \right]^{x_{ij}}$$

and

$$l(\boldsymbol{\theta}) = \sum_{i=1}^{n} \sum_{j=1}^{m} x_{ij} \log \int_{B_j} f(z; \boldsymbol{\theta}) \, \mathrm{d}z. \tag{2}$$

Write the MML estimator for $\boldsymbol{\theta}^0$ that is obtained from $n$ $z$-scores as $\hat{\boldsymbol{\theta}}_n$. We can define $\hat{\boldsymbol{\theta}}_n$ as a suitable root of the score equation $\nabla l = \mathbf{0}$, where $\nabla$ is the gradient operator and $\mathbf{0}$ is the zero vector.

The marginal likelihood function is simply an approximation to the likelihood that is constructed under an assumption of independence between the observations $\boldsymbol{X}_i$ (cf. Varin (2008)). In light of not knowing what the true dependence structure between the observations is, the marginal likelihood function can be seen as a quasi-likelihood construction in sense of White (1982); see also White (1994) and Spokoiny & Dickhaus (2015, Sec. 2.10). The purpose of a quasi-likelihood construction is to make use of an approximation that is close enough to the true data generative process so that meaningful inference can be drawn. Here its use is to avoid the need to declare an explicit model for potential correlation structures between the observations.

### 3.1 Maximum Marginal Likelihood Estimation

In order to compute $\hat{\boldsymbol{\theta}}_n$, we can utilize the EM algorithm of McLachlan & Jones (1988) for truncated and binned data. Suppose that we observe a realization $\boldsymbol{x}_i$ for each datum $\boldsymbol{X}_i$. Further, let $n_j = \sum_{i=1}^{n} x_{ij}$, for each $j \in [m]$. Define $\boldsymbol{\theta}^{(0)}$ to be some initial value of the EM algorithm and denote the value of $\boldsymbol{\theta}^{(r)}$ after the $r$th iteration by $\boldsymbol{\theta}^{(r)\top} = \left( \pi_0^{(r)}, \mu_0^{(r)}, \sigma_0^{2(r)}, \mu_1^{(r)}, \sigma_1^{2(r)} \right)$. Without going into the details of its derivation, the EM algorithm proceeds as follows.

On the $(r+1)$th E-step (expectation-step), compute $\alpha_{jk}^{(r+1)}$, $\beta_{jk}^{(r+1)}$, and $\gamma_{jk}^{(r+1)}$ for each $j \in [m]$ and $k \in \{0,1\}$, where

$$\alpha_{jk}^{(r+1)} = \frac{\pi_k^{(r)} \int_{B_j} \phi\left(z; \mu_k^{(r)}, \sigma_k^{2(r)}\right) \mathrm{d}z}{\int_{B_j} f\left(z; \boldsymbol{\theta}^{(r)}\right) \mathrm{d}z}, \tag{3}$$

$$\beta_{jk}^{(r+1)} = \frac{\pi_k^{(r)} \delta_{ik}^{(r+1)}}{\int_{B_j} f\left(z; \boldsymbol{\theta}^{(r)}\right) \mathrm{d}z}, \tag{4}$$

and

$$\gamma_{jk}^{(r+1)} = \frac{\pi_k^{(r)} \kappa_{ik}^{(r+1)}}{\int_{B_j} f\left(z; \boldsymbol{\theta}^{(r)}\right) \mathrm{d}z}. \tag{5}$$

Here

$$\delta_{jk}^{(r+1)} = \mu_k^{(r)} \int_{B_j} \phi\left(z; \mu_k^{(r)}, \sigma_k^{2(r)}\right) \mathrm{d}z - \sigma_k^{2(r)} v_{jk}^{(r+1)}$$

and

$$\kappa_{jk}^{(r+1)} = \sigma_k^{2(r)} \left[ \int_{B_j} \phi\left(z; \mu_k^{(r)}, \sigma_k^{2(r)}\right) \mathrm{d}z + \left(2\mu_k^{(r+1)} - \mu_k^{(r)}\right) v_{jk}^{(r+1)} - \omega_{jk}^{(r+1)} \right]$$
$$+ \left[2\mu_k^{(r+1)} - \mu_k^{(r)}\right]^2 v_{jk}^{(r+1)},$$

where

$$v_{jk}^{(r+1)} = \phi\left(b_j; \mu_k^{(r)}, \sigma_k^{2(r)}\right) - \phi\left(b_{j-1}; \mu_k^{(r)}, \sigma_k^{2(r)}\right),$$

$$\omega_{jk}^{(r+1)} = b_j \phi\left(b_j; \mu_k^{(r)}, \sigma_k^{2(r)}\right) - b_{j-1} \phi\left(b_{j-1}; \mu_k^{(r)}, \sigma_k^{2(r)}\right),$$

and $b_m = \infty$. Then, on the $(r+1)$th M-step (maximization-step), compute

$$\pi_k^{(r+1)} = n^{-1} \sum_{j=1}^{m} n_j \alpha_{jk}^{(r+1)}, \tag{6}$$

$$\mu_k^{(r+1)} = \left[\sum_{j=1}^{m} n_j \alpha_{jk}^{(r+1)}\right]^{-1} \sum_{j=1}^{m} n_j \beta_{jk}^{(r+1)}, \tag{7}$$

and

$$\sigma_k^{2(r+1)} = \left[\sum_{j=1}^{m} n_j \alpha_{jk}^{(r+1)}\right]^{-1} \sum_{j=1}^{m} n_j \gamma_{jk}^{(r+1)}, \tag{8}$$

for each $k \in \{0, 1\}$. The E-step and M-steps are repeated until some predetermined stopping criterion is met; see Lange (2013, Sec. 11.5) regarding stopping criteria. Upon stopping, the final iterate of the EM algorithm is declared the MML estimate $\hat{\boldsymbol{\theta}}$.

Since the algorithm composing of updates (3)–(8) constitutes an EM algorithm under the strict definition of Dempster et al. (1977) (see also McLachlan & Krishnan, 2008, Sec. 1.5), the usual properties of the EM algorithm, as proved by Wu (1983), are conferred upon it. That is, starting from some initial value $\boldsymbol{\theta}^{(0)}$, if we let $\boldsymbol{\theta}^{(\infty)} = \lim_{r \to \infty} \boldsymbol{\theta}^{(r)}$ be the limit point of the EM algorithm, then $\boldsymbol{\theta}^{(\infty)}$ is a stationary point of the log-marginal likelihood (2) and the sequence $l\left(\boldsymbol{\theta}^{(r)}\right)$ is monotonically increasing in $r$. See McLachlan & Krishnan (2008, Ch. 3) for details regarding the properties of EM algorithms. We note that the EM algorithm given above is that which is implemented in the **mixdist** package.

### 3.2 Consistency of the Estimator

As discussed in Bickel & Doksum (2001, Ch. 5), one of the most important properties of any large-sample estimator is that it is consistent (i.e. it converges to something meaningful as more data are obtained). We note that if one observes the data $\boldsymbol{X}_i$ and not $P_i$ or $Z_i$, for $i \in [n]$, then we can write the individual log-mass for each $\boldsymbol{X}_i$, given fixed bins $B_j$, as

$$\log \mathbb{P}\left(\boldsymbol{X}_i = \boldsymbol{x}; \boldsymbol{\theta}\right) = \prod_{j=1}^{m} \left[\int_{B_j} \pi_0 \phi\left(z; \mu_0, \sigma_0^2\right) + \pi_1 \phi\left(z; \mu_1, \sigma_1^2\right) \mathrm{d}z\right]^{x_{ij}}. \tag{9}$$

Substitution of (9) into (2) yields the log-marginal likelihood

$$l\left(\boldsymbol{\theta}\right) = \sum_{i=1}^{n} \log \mathbb{P}\left(\boldsymbol{X}_i = \boldsymbol{x}; \boldsymbol{\theta}\right).$$

Under mild assumptions regarding the dependence structure of the data $\boldsymbol{X}_1, \ldots, \boldsymbol{X}_n$, we can establish the consistency of the MML estimator $\hat{\boldsymbol{\theta}}_n$ via Theorem 5.14 of van der Vaart (1998). The result is as follows, and the proof can be found in the Appendix.

**Proposition 1.** *Assume that $\boldsymbol{X}_1, \boldsymbol{X}_2, \ldots, \boldsymbol{X}_n$ is an identical and strongly-dependent random sequence. Let $-\infty < m < M < \infty$, $0 < s < S < \infty$, and*

$$\Theta = \left\{\boldsymbol{\theta} : \pi_0 > 0, \pi_1 > 0, \pi_0 + \pi_1 = 1, (\mu_0, \mu_1) \in [m, M]^2, (\sigma_0^2, \sigma_1^2) \in [s, S]^2\right\}.$$

*If*

$$\Theta_0 = \left\{\boldsymbol{\theta}^0 \in \Theta : \mathbb{E}\log\mathbb{P}\left(\boldsymbol{X}_1 = \boldsymbol{x}; \boldsymbol{\theta}^0\right) = \sup_{\boldsymbol{\theta} \in \Theta} \mathbb{E}\log\mathbb{P}\left(\boldsymbol{X}_1 = \boldsymbol{x}; \boldsymbol{\theta}\right)\right\},$$

*then for every $\epsilon > 0$ and compact set $\mathbb{K} \subset \Theta$, we have*

$$\lim_{n \to \infty} \mathbb{P}\left(\sup_{\boldsymbol{\theta} \in \Theta_0} \left\|\hat{\boldsymbol{\theta}}_n - \boldsymbol{\theta}\right\| \geq \epsilon \text{ and } \hat{\boldsymbol{\theta}} \in \mathbb{K}\right) \to 0.$$

We note that an assumption that implies strong-mixing is $M$-dependence; see for example Bradley (2005). That is, if for each index $i$, the datum $\boldsymbol{X}_i$ is dependent on only $\boldsymbol{X}_j$, where $|i - j| \leq M < \infty$. This model is sufficient for many applied settings, such as genome studies and biological imaging.

A caveat to the application of the MML estimator is that one cannot always guarantee that $\hat{\boldsymbol{\theta}}$ is in fact the maximal value that is required in Proposition 1. This is because the EM algorithm is only guaranteed to converge to a local maximum of (2) (or a saddle-point that can easily be perturbed to continue onto a local maximum) and not the global maximum required by the theorems. This problem can be largely mitigated by using multiple runs of the EM algorithm from well-selected initial values. The topic of initialization of EM algorithms for mixture models is a complex one and discussions can be found in McLachlan (1988), Biernacki et al. (2003), Karlis & Xekalaki (2003), and Melnykov & Melnykov (2012).

## 3.3 Empirical Bayes-Based FDR Control

Upon estimation of the parameter vector $\boldsymbol{\theta}^0$ via the MML estimator $\hat{\boldsymbol{\theta}}$, we can follow the approach of McLachlan et al. (2006) in order to implement EB-based FDR control of the experiment; see also Nguyen et al. (2014). That is, consider the event $\{H_i \text{ is null } |Z_i = z_i\}$, for each $i \in [n]$. Via Bayes' rule and the MML estimator $\hat{\boldsymbol{\theta}}_n$, we can estimate the probability of the aforementioned event via the expression

$$\hat{\mathbb{P}}\left(H_i \text{ is null } |Z_i = z_i\right) = \frac{\hat{\pi}_0 \phi\left(z_i; \hat{\mu}_0, \hat{\sigma}_0^2\right)}{f\left(z_i; \hat{\boldsymbol{\theta}}_n\right)} = \tau\left(z_i; \hat{\boldsymbol{\theta}}\right). \qquad (10)$$

Using (10), we can then define the rejection rule

$$r\left(z_i; \hat{\boldsymbol{\theta}}_n, c\right) = \begin{cases} 1, & \text{if } \tau\left(z_i; \hat{\boldsymbol{\theta}}_n\right) \leq c \\ 0, & \text{otherwise,} \end{cases}$$

where $c \in [0, 1]$. Here $r\left(z_i; \hat{\boldsymbol{\theta}}_n, c\right) = 1$ if the null hypothesis of $H_i$ is rejected (i.e. $H_i$ is declared significant) and 0 otherwise.

Let the marginal FDR be defined as $m\text{FDR} = \mathbb{E}N_{01}/\mathbb{E}N_R$. We can estimate the $m$FDR of an experiment via the expression

$$\widehat{m\text{FDR}} = \frac{\sum_{i=1}^{n} \tau\left(z_i; \hat{\boldsymbol{\theta}}_n\right) \mathbb{I}\left(r\left(z_i; \hat{\boldsymbol{\theta}}_n, c\right) = 1\right)}{\sum_{i=1}^{n} \mathbb{I}\left(r\left(z_i; \hat{\boldsymbol{\theta}}_n, c\right) = 1\right)}, \quad (11)$$

which we can prove to converge to the $m$FDR in probability, under $M$-dependence (cf. Nguyen et al., 2014, Thm. 1). Subsequently, we can also demonstrate that for large $n$, the $m$FDR approaches the FDR (cf. Nguyen et al., 2014, Thm. 2).

Notice that $m\text{FDR} = m\text{FDR}(c)$ is a function of the threshold $c$. Using the thresholding value, we can approximately control the FDR at any desired level $\beta$ by setting the threshold $c$ using the rule

$$c_\beta = \arg\max\left\{c \in [0,1] : \widehat{m\text{FDR}}(c) \leq \beta\right\}. \quad (12)$$

### 3.4 Choosing the Binning Scheme

Thus far in discussing the binned estimation of the $z$-score distribution $f$, we have assumed that the bin cutoffs $b_1, ..., b_{m-1}$ are predetermined. However, the choice of a binning scheme is non-trivial.

A simple approach to the choice of binning scheme is to use the techniques underlying optimal histogram smoothing on the finite $z$-scores; see for example Wasserman (2006, Sec. 6.2). In **R**, there are several optimal histogram smoothing techniques that are deployed in the default **hist** function. These include the methods of Sturges (1926), Scott (1979), and Freedman & Diaconis (1981).

Under the methods of Sturges (1926), Scott (1979), and Freedman & Diaconis (1981), the number of bins is taken to be $m = \lceil \log_2 n \rceil + 1$,

$$m = \lceil (\text{Range}/h) \rceil \text{ with } h = 2 \times \text{IQR}/n^{1/3},$$

and

$$m = \lceil (\text{Range}/h) \rceil \text{ with } h = 3.5 \times s/n^{1/3},$$

respectively. Here, $\lceil \cdot \rceil$ is the ceiling operator, and Range, IQR, and $s$ are the sample range, interquartile range, and standard deviation, respectively. We compare the effectiveness of each of the binning approaches in the next section.

## 4 Assessment of the Binned Estimator

### 4.1 Accuracy of $z$-score Distribution

We first repeat the experiment from Section 2.3, except instead of ML estimation via the **normalmixEM2comp** function from the package **mixtools**, we conduct MML estimation via the **mix** function from the package **mixdist**. The results from the experiment, using binning schemes obtained via the histogram binning techniques of Sturges (1926), Scott (1979), and Freedman & Diaconis (1981) are reported in Table 3.

The first set of rows of Table (3) reports the MML estimation results when no encodings of testing data are implemented. We observe that the MML estimates appear to be accurate and demonstrate no statistically significant deviation away from the generative parameter elements of the model. The accuracy of the MML estimator appears to be robust to the choice among the three assessed binning schemes. This empirical result supports the theoretical conclusions of Proposition 1.

We note that there is only one set of table rows where we do not observe the uniform accuracy of the MML estimator, across the binning schemes that are applied. Under 8-bits $p$-type encoding, we observe that only the Sturges-binned MML estimator yielded accurate estimates of the generative parameter elements. Both the Freedman-Diaconis (FD) and Scott-binned estimators re-sulted in significantly inaccurate estimates of the null proportion and alternative mean and variance parameters.

Table 3: Mean and standard errors (SE) for the averages of 100 MML estimates of $\boldsymbol{\theta}$ based on $p$-type and $T$-type integer encodings of testing data. Here, FD denotes Freedman-Diaconis.

| Encoding | Binning | $\hat{\pi}_0$ Mean | SE | $\hat{\mu}_0$ Mean | SE | $\hat{\sigma}_0^2$ Mean | SE | $\hat{\mu}_1$ Mean | SE | $\hat{\sigma}_1^2$ Mean | SE |
|---|---|---|---|---|---|---|---|---|---|---|---|
| None | FD | 8.00E-01 | 1.53E-03 | 2.09E-04 | 4.07E-03 | 1.00E+00 | 2.95E-03 | 2.00E+00 | 5.17E-03 | 1.00E+00 | 5.35E-03 |
|  | Scott | 8.00E-01 | 1.53E-03 | -4.06E-04 | 4.37E-03 | 1.00E+00 | 3.53E-03 | 2.00E+00 | 4.99E-03 | 9.99E-01 | 5.50E-03 |
|  | Sturges | 8.00E-01 | 1.65E-03 | 1.21E-04 | 4.44E-03 | 1.00E+00 | 3.59E-03 | 2.00E+00 | 5.15E-03 | 1.00E+00 | 6.63E-03 |
| 8-bits $p$-type | FD | 7.12E-01 | 6.54E-02 | 1.91E-02 | 3.78E-02 | 1.00E+00 | 2.75E-02 | 1.48E+00 | 3.30E-01 | 1.50E+00 | 6.14E-02 |
|  | Scott | 7.52E-01 | 9.13E-02 | 7.61E-03 | 4.62E-02 | 1.01E+00 | 2.83E-02 | 1.77E+00 | 4.07E-01 | 1.26E+00 | 1.33E-01 |
|  | Sturges | 7.99E-01 | 2.55E-03 | -1.51E-03 | 6.94E-03 | 1.00E+00 | 4.89E-03 | 1.98E+00 | 7.62E-03 | 9.82E-01 | 1.25E-02 |
| 9-bits $p$-type | FD | 7.91E-01 | 5.55E-02 | 6.06E-03 | 3.42E-02 | 1.01E+00 | 1.68E-02 | 1.96E+00 | 2.21E-01 | 1.08E+00 | 8.60E-02 |
|  | Scott | 7.97E-01 | 4.37E-02 | 6.90E-03 | 2.69E-02 | 1.01E+00 | 1.30E-02 | 1.99E+00 | 1.61E-01 | 1.06E+00 | 5.55E-02 |
|  | Sturges | 8.05E-01 | 3.11E-03 | 1.38E-02 | 8.41E-03 | 1.01E+00 | 6.16E-03 | 2.03E+00 | 7.04E-03 | 1.06E+00 | 2.05E-02 |
| 16-bits $p$-type | FD | 8.00E-01 | 1.45E-03 | 1.66E-03 | 4.82E-03 | 1.00E+00 | 3.27E-03 | 2.00E+00 | 5.28E-03 | 1.01E+00 | 7.49E-03 |
|  | Scott | 8.00E-01 | 1.40E-03 | 1.24E-03 | 3.89E-03 | 1.00E+00 | 3.19E-03 | 2.00E+00 | 5.68E-03 | 1.01E+00 | 5.95E-03 |
|  | Sturges | 7.98E-01 | 1.16E-02 | -6.07E-04 | 1.42E-02 | 1.00E+00 | 5.69E-03 | 1.99E+00 | 5.35E-02 | 1.01E+00 | 1.96E-02 |
| 17-bits $p$-type | FD | 8.00E-01 | 4.30E-03 | -4.92E-04 | 6.50E-03 | 1.00E+00 | 3.78E-03 | 2.00E+00 | 1.94E-02 | 1.00E+00 | 8.66E-03 |
|  | Scott | 7.99E-01 | 7.95E-03 | -7.36E-04 | 1.02E-02 | 1.00E+00 | 4.55E-03 | 1.99E+00 | 3.73E-02 | 1.01E+00 | 1.31E-02 |
|  | Sturges | 8.00E-01 | 1.43E-03 | 5.46E-04 | 4.30E-03 | 1.00E+00 | 3.24E-03 | 2.00E+00 | 5.43E-03 | 1.00E+00 | 6.69E-03 |
| 7-bits $T$-type | FD | 7.98E-01 | 1.40E-02 | -3.99E-03 | 1.56E-02 | 9.98E-01 | 6.54E-03 | 1.99E+00 | 5.82E-03 | 1.00E+00 | 2.09E-02 |
|  | Scott | 7.98E-01 | 7.10E-03 | -3.71E-03 | 9.70E-03 | 9.99E-01 | 6.62E-03 | 2.00E+00 | 2.82E-02 | 1.00E+00 | 1.20E-02 |
|  | Sturges | 7.97E-01 | 1.57E-02 | -1.19E-02 | 1.70E-02 | 9.96E-01 | 1.13E-02 | 2.00E+00 | 6.73E-02 | 9.96E-01 | 2.42E-02 |
| 8-bits $T$-type | FD | 8.00E-01 | 3.55E-03 | -1.15E-03 | 5.31E-03 | 1.00E+00 | 3.92E-03 | 2.00E+00 | 1.67E-02 | 1.00E+00 | 7.06E-03 |
|  | Scott | 8.00E-01 | 1.39E-03 | -9.95E-04 | 3.79E-03 | 1.00E+00 | 4.12E-03 | 2.00E+00 | 4.50E-03 | 9.99E-01 | 5.76E-03 |
|  | Sturges | 7.99E-01 | 5.03E-03 | -4.47E-03 | 6.95E-03 | 9.98E-01 | 5.75E-03 | 2.00E+00 | 2.21E-02 | 9.98E-01 | 9.06E-03 |
| 15-bits $T$-type | FD | 8.00E-01 | 1.48E-03 | 7.81E-04 | 4.27E-03 | 1.00E+00 | 3.60E-03 | 2.00E+00 | 5.28E-03 | 1.00E+00 | 6.22E-03 |
|  | Scott | 8.00E-01 | 2.31E-03 | -3.40E-04 | 4.50E-03 | 1.00E+00 | 3.18E-03 | 2.00E+00 | 1.08E-02 | 9.99E-01 | 6.85E-03 |
|  | Sturges | 8.00E-01 | 1.49E-03 | 9.63E-05 | 4.54E-03 | 1.00E+00 | 3.16E-03 | 2.00E+00 | 5.65E-03 | 9.99E-01 | 6.27E-03 |
| 16-bits $T$-type | FD | 8.00E-01 | 1.93E-03 | -3.00E-04 | 5.13E-03 | 1.00E+00 | 3.29E-03 | 2.00E+00 | 8.88E-03 | 1.00E+00 | 7.04E-03 |
|  | Scott | 8.00E-01 | 3.29E-03 | -2.18E-04 | 6.09E-03 | 1.00E+00 | 3.55E-03 | 2.00E+00 | 1.56E-02 | 1.00E+00 | 8.44E-03 |
|  | Sturges | 8.00E-01 | 1.31E-03 | -4.17E-05 | 4.39E-03 | 1.00E+00 | 3.13E-03 | 2.00E+00 | 4.76E-03 | 1.00E+00 | 6.12E-03 |

Upon inspection, we found that the reason for the inaccuracy may be due to the fact that the FD and Scott binning methods yielded too many bins, that are of uniform width in the space of the $z$-scores. The $p$-type encodings generates uniform width rounding of data in the $p$-value space, which when converted to $z$-scores, can sometimes leave FD and Scott-type bins empty. This in turn causes the EM algorithm to fit the idiosyncratic nature of these empty bin patterns, that leads to overfitting and biased estimation. This problem diminishes as $\gamma$ increases, since there is more overlap between the encoded $p$-values and the FD and Scott-type bins, which leads to fewer numbers of empty bins, and thus less overfitting.

The Sturges binning mitigates against this empty bins problem by having much larger bin sizes than the other two assessed methods. We also note that the Sturges binning leads to faster EM algorithm runtimes due to the fact that fewer numerical integrals are required in the E-step, as described in Section 3.1. Since we do not observe any benefits from using FD or Scott-type binning in cases where all three methods yielded accurate estimates, we shall henceforth only consider the use of Sturges bins.

## 4.2 FDR Control Experiment

We perform a set of five numerical simulation scenarios, in order to assess the performance of the EB-based FDR control rule that is described in Section 3.3. These studies are denoted S1–S5, and will be described in the sequel.

In each of the scenarios, we generate $n = 10^6$ test statistics $T_1, \ldots, T_n$, with proportion $\pi_0 = 0.8$ that $H_i$ is null ($i \in [n]$). The generative distribution of $T_i$ given $H_i$ is null or alternative differs by the simulation study. However, under each studied scenario, the null hypothesis is assumed to be that $T_i$ is standard normal, and thus $p$-values are computed as $P_i = 1 - \Phi(T_i)$.

The $p$-values $P_1, \ldots, P_n$ then undergo the various valid $p$-type and $T$-type encodings that were previously considered. The EB-based FDR control method is then used to decide which of the hypotheses $H_i$ are significant, at the FDR control level $\beta \in \{0.05, 0.10\}$, based only on the encoded $p$-values. We compute the false discovery proportion (FDP) and true positive proportion (TPP) from the experiment as measures of performance of FDR control and testing power. The measures FDP and TPP are defined as FDP $= N_{01}/N_R$ and TPP $= N_{11}/N_1$, where $N_{11}$ is the number of false positives, $N_R$ is the number of rejected hypotheses (declared significantly alternative), $N_{11}$, is the number of true positives, and $N_1$ is the number of alternative hypotheses from the simulated experiment. For each simulation scenario, the experiment is repeated Reps $= 100$ times and the performance measurements are averaged over the repetitions.

For comparison, we also perform FDR control using the popular methods of Benjamini & Hochberg (1995) and Benjamini & Yekutieli (2001), which we denote as BH and BY, respectively. We also compare our EB-based FDR control to the EB-related FDR control technique of Storey (2002), which is commonly referred to as $q$-values. We implement the BH and BY methods via the base **R p.adjust** function. The $q$-values technique is implemented via the **qvalue** package (Storey et al., 2015).

### 4.3 Simulation Scenarios

In Scenario S1, we independently generate $T_i$ from a standard normal distribution, given that $H_i$ is null, and from a normal distribution with mean 2 and variance 1, otherwise. This scenario is identical to that which is studied Section 2.3. The scenario is ideal, in the sense that it fulfills the situation whereupon the hypotheses are generate test statistics that are IID and well-specified in the sense that the $p$-values $P_i$ are uniformly distributed under the null. All methods should adequately control the FDR in this case.

We consider hypothesis tests that generate dependent test statistics in Scenarios S2 and S3. In S2 two first-order autoregressive sequences of $n$ observations are generated. The null sequence is generated with mean coefficient 0, autoregressive coefficient 0.5, and normal errors with variances scaled so that the overall variance of the sequence is 1. The second chain is the same, except that the mean coefficient is 2 instead of zero. If $H_i$ is null, then $T_i$ is drawn from the first chain; otherwise $T_i$ is drawn from the second chain. See Amemiya (1985, Sec. 5.2) regarding autoregressive models. Scenario S3 is exactly the same as Scenario S2, except that the autoregressive coefficient is set to $-0.5$ instead of 0.5.

The two scenarios above are designed to test the performance of the methods when there are dependencies between the hypotheses. Since S1 only induces a positive correlation structure on the test statistics, all of the methods should be able to correctly control the FDR level in this case. In S2, negative correlations are induced between consecutive test statistics. Thus, there are no theoretical guarantees of the performance of BH in this case. The robustness of BH to positive correlation is proved in Benjamini & Yekutieli (2001) (see also Yekutieli, 2008). Robustness of BY to all forms of correlation is proved in Benjamini & Yekutieli (2001) and the performance of $q$-values under weak dependence is discussed in Storey & Tibshirani (2003).

In Scenario S4, we independently generate $T_i$ from a normal distribution with mean 1.5 and variance 1, given that $H_i$ is null, and from a normal distribution with mean 2.5 and variance 1, otherwise. This scenario is misspecified in the sense that the $p$-values $P_i$ are not computed under the correct null hypothesis.

Thus, the distribution of $P_i$ will not be uniform and thus the well-specified testing assumption of BH, BY, and $q$-values is not met. The scenario is still somewhat ideal for our EB-based method, since the $z$-scores distribution under the null is a normal distribution.

Lastly, in Scenario S5, we independently generate $T_i$ from a Student $t$ distribution with mean 0 and variance 1 and degrees of freedom 25, given that $H_i$, and from a Student $t$ distribution with mean 2 and variance 1 and degrees of freedom 25, otherwise. This scenario is also misspecified in the sense that the the $p$-values $P_i$ are not computed under the correct null hypothesis. It is also not ideal for our EB-based method, since the distribution of $z$-scores under the null is not normal. Thus, there are no performance guarantees for any of the assessed methods in this case.

## 4.4 Results

The results for Scenarios S1–S5 are reported in Tables 4–8, respectively. Before specifically covering any of the tables in detail, we shall make some general observations. Firstly, in terms of power (i.e. TPP), the FDR control methods follow the order: BY, EB, BH, and $q$-values, from least to most powerful. Similarly, with respect to conservatism of their FDR control (i.e. how much smaller FPP is to the nominal value $\beta$), we observe the same order: BY, EB, BH, and $q$-values, from most conservative to least. In fact, across the three well-specified testing scenarios (S1–S3), we observe that EB, BH, and BY were all conservative. These three initial observations were uniform across the different encoding methods.

Next, we observe that $q$-values can often result in anti-conservative control of the FDR (i.e. FDP consistently exceeding the nominal value $\beta$) in many scenarios and encoding types. For example in S1, we observe that $q$-values is

Table 4: Average FDP and TPP results (Reps = 100) for Scenario S1.

| Encoding | Method | FDP $\beta=0.05$ | FDP $\beta=0.10$ | TPP $\beta=0.05$ | TPP $\beta=0.10$ |
|---|---|---|---|---|---|
| None | EB | 2.46E-02 | 4.66E-02 | 1.22E-01 | 2.14E-01 |
|  | BH | 3.99E-02 | 7.96E-02 | 1.89E-01 | 3.25E-01 |
|  | BY | 3.19E-03 | 6.63E-03 | 1.12E-02 | 2.84E-02 |
|  | $q$-values | 5.03E-02 | 1.00E-01 | 2.28E-01 | 3.81E-01 |
| 8-bits $p$-type | EB | 4.01E-02 | 4.11E-02 | 1.90E-01 | 1.93E-01 |
|  | BH | 3.98E-02 | 9.64E-02 | 1.88E-01 | 3.70E-01 |
|  | BY | 3.98E-02 | 3.98E-02 | 1.88E-01 | 1.88E-01 |
|  | $q$-values | 7.23E-02 | 1.13E-01 | 3.01E-01 | 4.12E-01 |
| 9-bits $p$-type | EB | 2.88E-02 | 4.33E-02 | 1.42E-01 | 2.00E-01 |
|  | BH | 4.92E-02 | 8.29E-02 | 2.25E-01 | 3.34E-01 |
|  | BY | 2.75E-02 | 2.75E-02 | 1.36E-01 | 1.36E-01 |
|  | $q$-values | 4.94E-02 | 1.02E-01 | 2.26E-01 | 3.85E-01 |
| 16-bits $p$-type | EB | 2.47E-02 | 4.63E-02 | 1.22E-01 | 2.13E-01 |
|  | BH | 3.99E-02 | 7.98E-02 | 1.88E-01 | 3.25E-01 |
|  | BY | 3.32E-03 | 6.72E-03 | 1.67E-02 | 3.08E-02 |
|  | $q$-values | 4.96E-02 | 9.98E-02 | 2.26E-01 | 3.80E-01 |
| 17-bits $p$-type | EB | 2.50E-02 | 4.64E-02 | 1.21E-01 | 2.13E-01 |
|  | BH | 4.02E-02 | 7.95E-02 | 1.88E-01 | 3.25E-01 |
|  | BY | 2.73E-03 | 7.07E-03 | 1.31E-02 | 3.01E-02 |
|  | $q$-values | 5.06E-02 | 1.00E-01 | 2.28E-01 | 3.80E-01 |
| 7-bits $T$-type | EB | 2.70E-02 | 5.10E-02 | 1.32E-01 | 2.27E-01 |
|  | BH | 4.21E-02 | 8.32E-02 | 1.95E-01 | 3.33E-01 |
|  | BY | 3.38E-03 | 7.02E-03 | 1.27E-02 | 3.08E-02 |
|  | $q$-values | 5.27E-02 | 1.04E-01 | 2.34E-01 | 3.88E-01 |
| 8-bits $T$-type | EB | 2.54E-02 | 4.78E-02 | 1.27E-01 | 2.20E-01 |
|  | BH | 4.01E-02 | 8.14E-02 | 1.90E-01 | 3.29E-01 |
|  | BY | 3.24E-03 | 6.63E-03 | 1.18E-02 | 2.97E-02 |
|  | $q$-values | 5.02E-02 | 1.02E-01 | 2.29E-01 | 3.84E-01 |
| 15-bits $T$-type | EB | 2.50E-02 | 4.72E-02 | 1.23E-01 | 2.15E-01 |
|  | BH | 4.08E-02 | 8.02E-02 | 1.89E-01 | 3.26E-01 |
|  | BY | 3.27E-03 | 6.92E-03 | 1.15E-02 | 2.89E-02 |
|  | $q$-values | 5.07E-02 | 9.99E-02 | 2.28E-01 | 3.81E-01 |
| 16-bits $T$-type | EB | 2.46E-02 | 4.66E-02 | 1.22E-01 | 2.14E-01 |
|  | BH | 4.05E-02 | 7.97E-02 | 1.88E-01 | 3.25E-01 |
|  | BY | 3.19E-03 | 6.60E-03 | 1.12E-02 | 2.87E-02 |
|  | $q$-values | 5.01E-02 | 1.00E-01 | 2.27E-01 | 3.80E-01 |

Table 5: Average FDP and TPP results (Reps = 100) for Scenario S2.

| Encoding | Method | FDP $\beta = 0.05$ | FDP $\beta = 0.10$ | TPP $\beta = 0.05$ | TPP $\beta = 0.10$ | Encoding | Method | FDP $\beta = 0.05$ | FDP $\beta = 0.10$ | TPP $\beta = 0.05$ | TPP $\beta = 0.10$ |
|---|---|---|---|---|---|---|---|---|---|---|---|
| None | EB | 2.44E-02 | 4.62E-02 | 1.23E-01 | 2.16E-01 | | | | | | |
| | BH | 3.96E-02 | 7.96E-02 | 1.90E-01 | 3.26E-01 | | | | | | |
| | BY | 2.99E-03 | 6.16E-03 | 1.16E-02 | 2.87E-02 | | | | | | |
| | $q$-values | 4.96E-02 | 1.00E-01 | 2.28E-01 | 3.81E-01 | | | | | | |
| 8-bits $p$-type | EB | 3.99E-02 | 3.99E-02 | 1.88E-01 | 1.88E-01 | 7-bits $T$-type | EB | 2.76E-02 | 5.12E-02 | 1.35E-01 | 2.31E-01 |
| | BH | 3.99E-02 | 9.65E-02 | 1.88E-01 | 3.69E-01 | | BH | 4.16E-02 | 8.24E-02 | 1.95E-01 | 3.33E-01 |
| | BY | 3.99E-02 | 3.99E-02 | 1.88E-01 | 1.88E-01 | | BY | 4.01E-03 | 6.94E-03 | 1.20E-02 | 3.04E-02 |
| | $q$-values | 7.26E-02 | 1.12E-01 | 3.02E-01 | 4.09E-01 | | $q$-values | 5.17E-02 | 1.02E-01 | 2.34E-01 | 3.88E-01 |
| 9-bits $p$-type | EB | 2.81E-02 | 4.23E-02 | 1.37E-01 | 1.97E-01 | 8-bits $T$-type | EB | 2.57E-02 | 4.82E-02 | 1.26E-01 | 2.20E-01 |
| | BH | 4.92E-02 | 8.23E-02 | 2.25E-01 | 3.33E-01 | | BH | 4.08E-02 | 8.10E-02 | 1.91E-01 | 3.28E-01 |
| | BY | 2.81E-02 | 2.81E-02 | 1.37E-01 | 1.37E-01 | | BY | 2.96E-03 | 6.63E-03 | 1.22E-02 | 2.95E-02 |
| | $q$-values | 4.95E-02 | 1.02E-01 | 2.26E-01 | 3.85E-01 | | $q$-values | 5.09E-02 | 1.02E-01 | 2.30E-01 | 3.84E-01 |
| 16-bits $p$-type | EB | 2.45E-02 | 4.67E-02 | 1.23E-01 | 2.14E-01 | 15-bits $T$-type | EB | 2.48E-02 | 4.64E-02 | 1.21E-01 | 2.13E-01 |
| | BH | 4.00E-02 | 8.05E-02 | 1.89E-01 | 3.26E-01 | | BH | 3.99E-02 | 7.98E-02 | 1.88E-01 | 3.24E-01 |
| | BY | 3.96E-03 | 6.90E-03 | 1.74E-02 | 3.12E-02 | | BY | 3.25E-03 | 6.42E-03 | 1.14E-02 | 2.83E-02 |
| | $q$-values | 5.04E-02 | 1.00E-01 | 2.27E-01 | 3.80E-01 | | $q$-values | 4.99E-02 | 9.99E-02 | 2.27E-01 | 3.80E-01 |
| 17-bits $p$-type | EB | 2.42E-02 | 4.62E-02 | 1.21E-01 | 2.13E-01 | 16-bits $T$-type | EB | 2.48E-02 | 4.73E-02 | 1.22E-01 | 2.15E-01 |
| | BH | 4.00E-02 | 8.01E-02 | 1.88E-01 | 3.25E-01 | | BH | 4.03E-02 | 8.01E-02 | 1.88E-01 | 3.24E-01 |
| | BY | 4.63E-03 | 7.27E-03 | 1.31E-02 | 3.00E-02 | | BY | 3.41E-03 | 6.64E-03 | 1.14E-02 | 2.88E-02 |
| | $q$-values | 5.00E-02 | 1.00E-01 | 2.27E-01 | 3.80E-01 | | $q$-values | 5.07E-02 | 1.01E-01 | 2.27E-01 | 3.80E-01 |

Table 6: Average FDP and TPP results (Reps = 100) for Scenario S3.

| Encoding | Method | FDP $\beta = 0.05$ | FDP $\beta = 0.10$ | TPP $\beta = 0.05$ | TPP $\beta = 0.10$ | Encoding | Method | FDP $\beta = 0.05$ | FDP $\beta = 0.10$ | TPP $\beta = 0.05$ | TPP $\beta = 0.10$ |
|---|---|---|---|---|---|---|---|---|---|---|---|
| None | EB | 2.52E-02 | 4.71E-02 | 1.22E-01 | 2.14E-01 | | | | | | |
| | BH | 4.04E-02 | 7.98E-02 | 1.89E-01 | 3.25E-01 | | | | | | |
| | BY | 3.11E-03 | 7.10E-03 | 1.16E-02 | 2.88E-02 | | | | | | |
| | $q$-values | 5.06E-02 | 9.97E-02 | 2.28E-01 | 3.81E-01 | | | | | | |
| 8-bits $p$-type | EB | 4.02E-02 | 4.05E-02 | 1.88E-01 | 1.89E-01 | 7-bits $T$-type | EB | 2.64E-02 | 4.94E-02 | 1.29E-01 | 2.24E-01 |
| | BH | 4.02E-02 | 9.60E-02 | 1.88E-01 | 3.69E-01 | | BH | 4.13E-02 | 8.22E-02 | 1.94E-01 | 3.31E-01 |
| | BY | 4.02E-02 | 4.02E-02 | 1.88E-01 | 1.88E-01 | | BY | 3.43E-03 | 7.39E-03 | 1.23E-02 | 3.05E-02 |
| | $q$-values | 7.21E-02 | 1.12E-01 | 3.02E-01 | 4.10E-01 | | $q$-values | 5.20E-02 | 1.03E-01 | 2.34E-01 | 3.88E-01 |
| 9-bits $p$-type | EB | 2.90E-02 | 4.51E-02 | 1.41E-01 | 2.05E-01 | 8-bits $T$-type | EB | 2.57E-02 | 4.80E-02 | 1.26E-01 | 2.20E-01 |
| | BH | 4.97E-02 | 8.20E-02 | 2.25E-01 | 3.31E-01 | | BH | 4.03E-02 | 8.08E-02 | 1.90E-01 | 3.28E-01 |
| | BY | 2.78E-02 | 2.78E-02 | 1.36E-01 | 1.36E-01 | | BY | 3.91E-03 | 6.72E-03 | 1.18E-02 | 2.93E-02 |
| | $q$-values | 4.98E-02 | 1.01E-01 | 2.26E-01 | 3.84E-01 | | $q$-values | 5.06E-02 | 1.01E-01 | 2.30E-01 | 3.83E-01 |
| 16-bits $p$-type | EB | 2.48E-02 | 4.60E-02 | 1.22E-01 | 2.14E-01 | 15-bits $T$-type | EB | 2.51E-02 | 4.64E-02 | 1.23E-01 | 2.15E-01 |
| | BH | 3.99E-02 | 7.97E-02 | 1.89E-01 | 3.25E-01 | | BH | 3.96E-02 | 7.95E-02 | 1.88E-01 | 3.24E-01 |
| | BY | 4.20E-03 | 6.93E-03 | 1.71E-02 | 3.09E-02 | | BY | 3.42E-03 | 6.61E-03 | 1.14E-02 | 2.85E-02 |
| | $q$-values | 4.98E-02 | 9.98E-02 | 2.28E-01 | 3.80E-01 | | $q$-values | 4.94E-02 | 9.99E-02 | 2.27E-01 | 3.80E-01 |
| 17-bits $p$-type | EB | 2.50E-02 | 4.66E-02 | 1.22E-01 | 2.14E-01 | 16-bits $T$-type | EB | 2.42E-02 | 4.65E-02 | 1.22E-01 | 2.14E-01 |
| | BH | 4.00E-02 | 7.98E-02 | 1.88E-01 | 3.24E-01 | | BH | 4.00E-02 | 8.02E-02 | 1.89E-01 | 3.25E-01 |
| | BY | 3.42E-03 | 7.05E-03 | 1.28E-02 | 2.97E-02 | | BY | 3.10E-03 | 6.59E-03 | 1.15E-02 | 2.90E-02 |
| | $q$-values | 4.99E-02 | 1.00E-01 | 2.27E-01 | 3.80E-01 | | $q$-values | 5.02E-02 | 1.00E-01 | 2.27E-01 | 3.80E-01 |

Table 7: Average FDP and TPP results (Reps = 100) for Scenario S4.

| Encoding | Method | FDP $\beta = 0.05$ | FDP $\beta = 0.10$ | TPP $\beta = 0.05$ | TPP $\beta = 0.10$ | Encoding | Method | FDP $\beta = 0.05$ | FDP $\beta = 0.10$ | TPP $\beta = 0.05$ | TPP $\beta = 0.10$ |
|---|---|---|---|---|---|---|---|---|---|---|---|
| None | EB | 2.48E-02 | 4.63E-02 | 1.22E-01 | 2.14E-01 | | | | | | |
| | BH | 1.48E-01 | 2.47E-01 | 4.89E-01 | 6.50E-01 | | | | | | |
| | BY | 2.17E-02 | 3.69E-02 | 1.07E-01 | 1.77E-01 | | | | | | |
| | $q$-values | 3.74E-01 | 5.69E-01 | 7.93E-01 | 9.33E-01 | | | | | | |
| 8-bits $p$-type | EB | 8.95E-02 | 8.95E-02 | 3.50E-01 | 3.50E-01 | 7-bits $T$-type | EB | 2.56E-02 | 4.80E-02 | 1.27E-01 | 2.20E-01 |
| | BH | 1.50E-01 | 2.49E-01 | 4.92E-01 | 6.53E-01 | | BH | 1.51E-01 | 2.52E-01 | 4.96E-01 | 6.57E-01 |
| | BY | 8.95E-02 | 8.95E-02 | 3.50E-01 | 3.50E-01 | | BY | 2.22E-02 | 3.81E-02 | 1.10E-01 | 1.82E-01 |
| | $q$-values | 3.80E-01 | 5.72E-01 | 7.99E-01 | 9.34E-01 | | $q$-values | 3.80E-01 | 5.74E-01 | 7.98E-01 | 9.35E-01 |
| 9-bits $p$-type | EB | 6.46E-02 | 6.46E-02 | 2.74E-01 | 2.74E-01 | 8-bits $T$-type | EB | 2.54E-02 | 4.73E-02 | 1.23E-01 | 2.15E-01 |
| | BH | 1.61E-01 | 2.53E-01 | 5.13E-01 | 6.58E-01 | | BH | 1.50E-01 | 2.50E-01 | 4.91E-01 | 6.53E-01 |
| | BY | 6.46E-02 | 6.46E-02 | 2.74E-01 | 2.74E-01 | | BY | 2.22E-02 | 3.80E-02 | 1.08E-01 | 1.79E-01 |
| | $q$-values | 3.76E-01 | 5.69E-01 | 7.95E-01 | 9.33E-01 | | $q$-values | 3.78E-01 | 5.72E-01 | 7.96E-01 | 9.35E-01 |
| 16-bits $p$-type | EB | 2.51E-02 | 4.74E-02 | 1.25E-01 | 2.17E-01 | 15-bits $T$-type | EB | 2.53E-02 | 4.68E-02 | 1.21E-01 | 2.13E-01 |
| | BH | 1.48E-01 | 2.47E-01 | 4.89E-01 | 6.50E-01 | | BH | 1.49E-01 | 2.47E-01 | 4.88E-01 | 6.50E-01 |
| | BY | 2.17E-02 | 3.70E-02 | 1.10E-01 | 1.77E-01 | | BY | 2.20E-02 | 3.75E-02 | 1.07E-01 | 1.76E-01 |
| | $q$-values | 3.74E-01 | 5.69E-01 | 7.93E-01 | 9.33E-01 | | $q$-values | 3.74E-01 | 5.68E-01 | 7.93E-01 | 9.32E-01 |
| 17-bits $p$-type | EB | 2.45E-02 | 4.65E-02 | 1.22E-01 | 2.13E-01 | 16-bits $T$-type | EB | 2.45E-02 | 4.65E-02 | 1.23E-01 | 2.15E-01 |
| | BH | 1.49E-01 | 2.47E-01 | 4.90E-01 | 6.51E-01 | | BH | 1.48E-01 | 2.47E-01 | 4.89E-01 | 6.51E-01 |
| | BY | 2.19E-02 | 3.73E-02 | 1.09E-01 | 1.78E-01 | | BY | 2.13E-02 | 3.65E-02 | 1.07E-01 | 1.77E-01 |
| | $q$-values | 3.74E-01 | 5.68E-01 | 7.92E-01 | 9.32E-01 | | $q$-values | 3.75E-01 | 5.70E-01 | 7.94E-01 | 9.33E-01 |

Table 8: Average FDP and TPP results (Reps = 100) for Scenario S5.

| Encoding | Method | FDP $\beta = 0.05$ | FDP $\beta = 0.10$ | TPP $\beta = 0.05$ | TPP $\beta = 0.10$ | Encoding | Method | FDP $\beta = 0.05$ | FDP $\beta = 0.10$ | TPP $\beta = 0.05$ | TPP $\beta = 0.10$ |
|---|---|---|---|---|---|---|---|---|---|---|---|
| None | EB | 4.35E-02 | 6.15E-02 | 1.02E-01 | 1.85E-01 | | | | | | |
| | BH | 6.10E-02 | 9.62E-02 | 1.82E-01 | 3.20E-01 | | | | | | |
| | BY | 2.48E-02 | 2.80E-02 | 1.51E-02 | 3.09E-02 | | | | | | |
| | $q$-values | 7.14E-02 | 1.16E-01 | 2.26E-01 | 3.85E-01 | | | | | | |
| 8-bits $p$-type | EB | 1.03E-01 | 1.03E-01 | 3.46E-01 | 3.46E-01 | 7-bits $T$-type | EB | 4.55E-02 | 6.55E-02 | 1.13E-01 | 2.01E-01 |
| | BH | 1.56E-01 | 2.45E-01 | 4.93E-01 | 6.58E-01 | | BH | 6.25E-02 | 9.85E-02 | 1.88E-01 | 3.30E-01 |
| | BY | 1.03E-01 | 1.03E-01 | 3.46E-01 | 3.46E-01 | | BY | 2.39E-02 | 2.83E-02 | 1.56E-02 | 3.18E-02 |
| | $q$-values | 3.65E-01 | 5.62E-01 | 8.00E-01 | 9.33E-01 | | $q$-values | 7.28E-02 | 1.19E-01 | 2.33E-01 | 3.94E-01 |
| 9-bits $p$-type | EB | 8.23E-02 | 8.23E-02 | 2.70E-01 | 2.70E-01 | 8-bits $T$-type | EB | 4.65E-02 | 6.67E-02 | 1.17E-01 | 2.06E-01 |
| | BH | 1.66E-01 | 2.50E-01 | 5.15E-01 | 6.66E-01 | | BH | 6.25E-02 | 9.83E-02 | 1.88E-01 | 3.28E-01 |
| | BY | 8.23E-02 | 8.23E-02 | 2.70E-01 | 2.70E-01 | | BY | 2.47E-02 | 2.71E-02 | 1.57E-02 | 3.27E-02 |
| | $q$-values | 3.63E-01 | 5.59E-01 | 7.97E-01 | 9.32E-01 | | $q$-values | 7.31E-02 | 1.18E-01 | 2.32E-01 | 3.92E-01 |
| 16-bits $p$-type | EB | 3.97E-02 | 5.67E-02 | 8.55E-02 | 1.62E-01 | 15-bits $T$-type | EB | 4.46E-02 | 6.20E-02 | 1.02E-01 | 1.85E-01 |
| | BH | 1.55E-01 | 2.43E-01 | 4.92E-01 | 6.56E-01 | | BH | 6.10E-02 | 9.59E-02 | 1.81E-01 | 3.20E-01 |
| | BY | 4.45E-02 | 5.86E-02 | 1.06E-01 | 1.72E-01 | | BY | 2.35E-02 | 2.85E-02 | 1.50E-02 | 3.06E-02 |
| | $q$-values | 3.61E-01 | 5.60E-01 | 7.97E-01 | 9.33E-01 | | $q$-values | 7.12E-02 | 1.15E-01 | 2.24E-01 | 3.84E-01 |
| 17-bits $p$-type | EB | 4.12E-02 | 5.73E-02 | 8.53E-02 | 1.63E-01 | 16-bits $T$-type | EB | 4.42E-02 | 6.14E-02 | 1.02E-01 | 1.85E-01 |
| | BH | 1.54E-01 | 2.42E-01 | 4.90E-01 | 6.55E-01 | | BH | 6.06E-02 | 9.55E-02 | 1.82E-01 | 3.21E-01 |
| | BY | 4.48E-02 | 5.88E-02 | 1.03E-01 | 1.70E-01 | | BY | 2.60E-02 | 2.96E-02 | 1.47E-02 | 3.03E-02 |
| | $q$-values | 3.60E-01 | 5.58E-01 | 7.95E-01 | 9.32E-01 | | $q$-values | 7.10E-02 | 1.16E-01 | 2.26E-01 | 3.86E-01 |

anti-conservative for both values of $\beta$ when we use 8-bits $p$-type encodings and for the $\beta = 0.05$ level when we use 7-bits $T$-type encodings. The same can be observed from the results of S2 and S3. This leads to our first recommendation from this study, which is that $q$-values should be avoided when data are compressed using 8-bits integers encoding.

In Scenario S4, we observe that BH and $q$-values are highly anti-conservative. Here applications of the two methods resulted in FDP values that greatly exceeded the nominal value of $\beta$, uniformly over the encoding methods. Both EB and BY were also anti-conservative when the data were compressed by either 8-bits or 9-bits $p$-type encodings, for control of the FDR at rate $\beta = 0.05$, although the exceedances were much less than those of the BH and $q$-values results. At the $\beta = 0.10$ level, both methods were conservative for the two previously mentioned encoding types. In all other encoding types, EB and BY were conservative. EB was less conservative and more powerful than BY in each of the cases where they both correctly controlled the FDR level, and thus should be preferred.

From Table 8, we observe that $q$-values were anti-conservative uniformly over encoding types and FDR control levels in Scenario S5. Furthermore, BH was also uniformly anti-conservative when used to control the FDR at $\beta = 0.05$. The BH method also yielded anti-conservative control of the FDR at $\beta = 0.10$, when the data were encoded using $p$-type encodings. Both EB and BY were equally anti-conservative for control of FDR at $\beta = 0.05$, when the data were encoded using 8-bits or 9-bits $p$-type encodings. However, the control at the $\beta = 0.10$ level from both methods for the two aforementioned encoding schemes were both equal and approximately at the correct rate. For all other encoding types, both EB and BY correctly controlled the FDR, for both levels of $\beta$. EB was more powerful in the $T$-type encodings, whereas BY was more powerful under $p$-type encodings. Thus, the better method depends on knowledge of which encoding type is used. However, we note that BY is only more powerful than EB by a small amount, under the $p$-type encodings, whereas EB can be more powerful than BY by an order of magnitude, under no encoding or $T$-type encodings.

From the results of Simulations S1–S5, we can conclude that the EB method can correctly control the FDR when the tests are well-specified, and are also somewhat robust to misspecification. Furthermore, EB along with BY are somewhat more robust to misspecification and data compression via integer encoding than the two other tested methods. We make the final observation that the $T$-type encodings tended to result in performance rates that were closer to those obtained from uncompressed data. Thus, when the choice is available, one should opt for $T$-type over $p$-type encodings, holding constant the bit rate of the compression.

## 5  Example Application

### 5.1  Description of Data

Correlations between the structural properties of brain regions, as measured over a sample of subjects, are being increasingly studied as a means of understanding neurological development (Li et al., 2013) and diseases (Seeley et al., 2009; Wheeler & Voineskos, 2014; Sharda et al., 2016). These correlation patterns, which are often referred to as structural covariance in the neuroimaging literature, are widely studied in humans (Alexander-Bloch et al., 2013; Evans, 2013), as well as in animal models such as mice (Pagani et al., 2016).

For our example application, we study neurological magnetic resonance imaging (MRI) data from a sample of 241 mice. The MRI sample of both female and male adult mice were obtained by taking the control data from a phenotyping study (Ellegood et al., 2015) in order to create a representative wildtype population with variability. All mice were scanned ex-vivo after perfusion with a gadolinium-based contrast agent, and all images were obtained at the same location (i.e. the Mouse Imaging Centre). Scanning was performed on a Varian 7T small animal MR scanner that was adapted for multiple mouse imaging.

The preparation and image acquisition followed a standard pipeline that is similar to the one described in Lerch et al. (2010). Specifically, a T2-weighted fast-spin echo sequence was used to produce whole-brain images that have an isotropic resolution of 56 micrometers. After images were acquired, the data were corrected for distortions and then registered together by deformation towards a common nonlinear average. The registration pipeline included corrections for nonuniformities that were induced by radio frequency inhomogeneities or gradient-related eddy currents (Sled et al., 1998). The registered images had a volume of $x \times y \times z = 225 \times 320 \times 152$ voxels, of which $n = 2818191$ voxels corresponded to neurological matter. The exported data were stored in the MINC format.

As an output, the registration process produces a set of Jacobian determinants that provide a measure of the extent in which a voxel from the average brain must expand or contract in order to match each of the individual brains of the sample. The Jacobian determinants field of each sample individual is thus a measure of local volume change. For further processing, the Jacobian determinants are log-transformed in order to reduce skewness.

## 5.2 Hypothesis Testing

Upon attainment of the sample of 241 Jacobian determinant fields from the registered mice brain MRIs, we can assess whether or not the local volume change at any particular voxel is correlated with some region of interest. To do so, we select a "seed" voxel within the region of interest and compute the voxelwise sample (Pearson) correlation between the log-transformed Jacobian determinant of the seed voxel and those at every other voxel in the sample of MRIs. This correlation measure can then be used as a measure of structural covariance of the region of interest and the rest of the brain. In the past, structural covariance methods have been used to draw inference regarding a broad array of phenomena such as cortical thickness (Lerch et al., 2006), and cortical maturation and development (Raznahan et al., 2011).

Thus at each of the $n = 2818191$ voxels we computed a correlation coefficient. Using the correlation coefficients, we conducted voxelwise tests of the null hypothesis that the true correlation between the log-transformed Jacobian determinants of the seed voxel and voxel $i \in [n]$ is zero versus the two-sided alternative. The $p$-values of each test were computed using the Fisher $z$-transformation and normal approximation (Fisher, 1915, 1921).

Using the seed voxel at spatial location $(x, y, z) = (125, 124, 64)$ – within the bed nucleus of the stria terminalis – we conducted the hypothesis tests, as described above. Histograms of the $p$-values and log-squared correlation coefficients can be found in Figure 1. We note that the histogram of the log-squared correlation coefficients omits 35856 voxels that had zero correlation with the seed voxel. Further note that a correlation of one yields a log-squared coefficient of $\approx -0.69$.

An inspection of Figure 1 reveals that the $p$-value distribution from the experiment deviates significantly from a uniform distribution. The magnitude of the deviation indicates that there may be a potentially large number of voxels that are strongly correlated with the seed voxel, and thus with the region of interest that the seed voxel represents. Using FDR control, we can attempt to identify these correlated voxels in a manner that limits the potential number of false discoveries that are made.

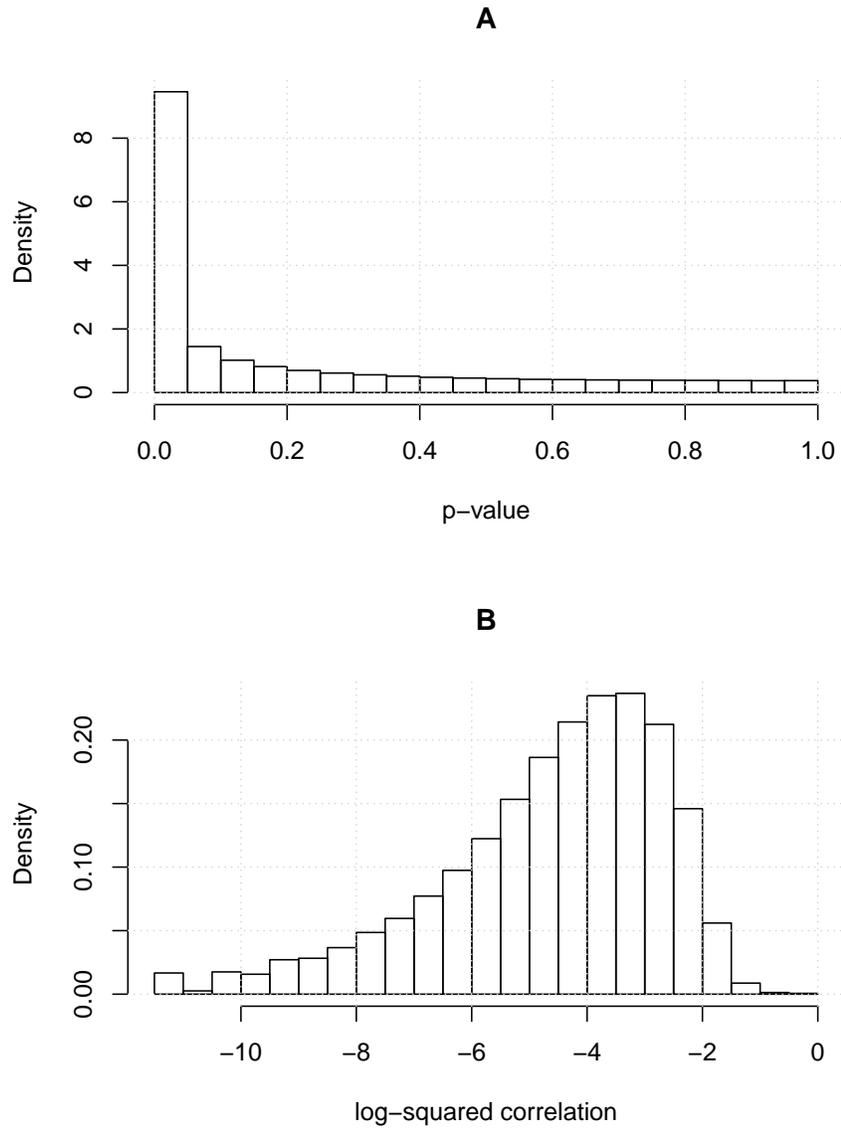

Figure 1: Histograms of $p$-values and log-squared correlation coefficients for the structural covariance experiment with seed voxel $(x, y, z) = (125, 124, 64)$ are presented in subplot A and B, respectively.

A closer inspection of the $p$-value distribution reveals that there are only 66249 discrete and unique numerical values that make up the sample. These discrete values include zero and one, making up 311575 and 6 voxels of the $p$-value sample, respectively. Our observations indicate that the data stored as signed or unsigned 16-bit integers, at some stage in processing pipeline. It is difficult to tell since there may have been multiple encodings of the data along the pipeline that has resulted in the final reported outputs. As such, from our earlier discussions, it would be prudent to apply our EB-based FDR control methodology, since it explicitly accounts for the encoded nature of the data. Furthermore, due to the mathematical approximation via the use of the Fisher $z$-transformation as well as the omission of other variables that may contribute to the analysis such as covariates describing the mice (e.g. gender and model strain), the null hypothesis that the population correlation is equal to zero is likely to be misspecified. From Section 4.3, we have observed that the EB-based method is effective in such a setting.

## 5.3 FDR Control

We firstly transform the $p$-values $p_i$ to the $z$-scores $p_i = \Phi^{-1}(1 - p_i)$, for each $i \in [n]$. A histogram of the $z$-scores that is obtained is presented in Figure 2. We note that the $z$-scores that are obtained from the 311581 with $p$-values equal to zero or one are omitted in this plot. There is a clear truncation of the histogram at the $z$-score value of 4.169 which corresponds to the smallest non-zero $p$-value of 1.53E-05.

Using the methods from Section 3, we fit the EB mixture model and obtain

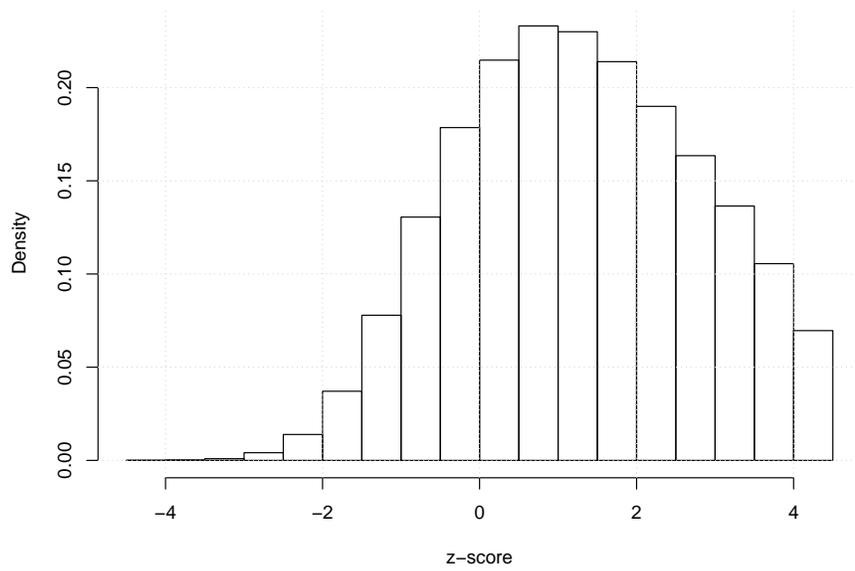

Figure 2: Histogram of $z$-values excluding those from voxels with $p$-values equal to zero or one.

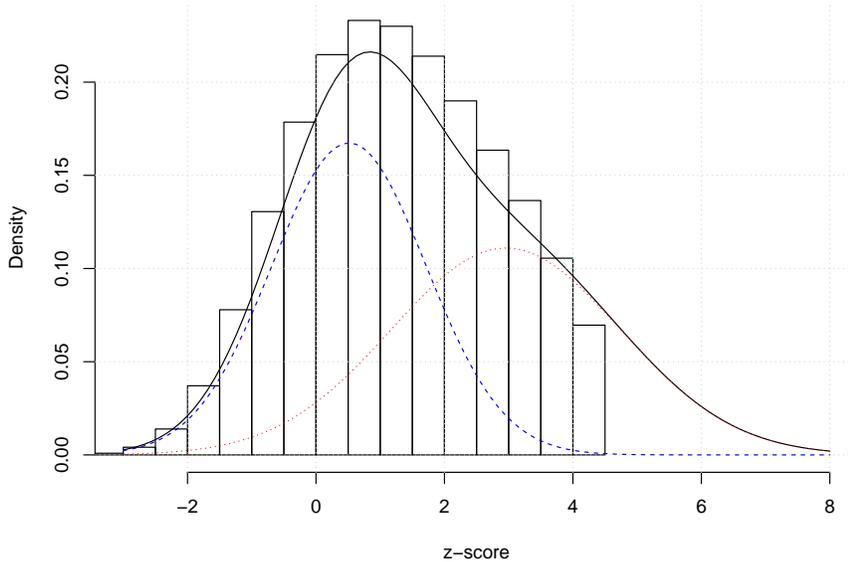

Figure 3: The functions $f\left(\cdot;\hat{\boldsymbol{\theta}}\right)$, $\hat{\pi}_0\hat{f}_0$, and $\hat{\pi}_1\hat{f}_1$ are plotted with solid, dashed, and dotted lines, respectively.

the parameter vector

$$\begin{aligned}\hat{\boldsymbol{\theta}}^\top &= \left(\hat{\pi}_0, \hat{\mu}_0, \hat{\sigma}_0^2, \hat{\mu}_1, \hat{\sigma}_1^2\right) \\ &= \left(0.5035, 0.5141, 1.200^2, 2.9568, 1.785^2\right),\end{aligned} \quad (13)$$

which corresponds to the mixture model,

$$f\left(z;\hat{\boldsymbol{\theta}}\right) = 0.5035\phi\left(z; 0.5141, 1.200^2\right) + 0.4965\phi\left(z; 2.9568, 1.785^2\right). \quad (14)$$

Let $\hat{f}_0(z) = \phi\left(z; 0.5141, 1.200^2\right)$ and $\hat{f}_1(z) = \phi\left(z; 2.9568, 1.785^2\right)$ be the estimates of $f_0$ and $f_1$, respectively. We visualize $f\left(\cdot;\hat{\boldsymbol{\theta}}\right)$, $\hat{\pi}_0\hat{f}_0$, and $\hat{\pi}_1\hat{f}_1$ together in Figure 3.

Upon inspection of Figure 3, we observe that mixture model (14) provides a good fit to the suggested curvature of the histogram. The estimated parameter vector from (13) indicates that the null distribution is significantly shifted to the right. This may be due to a combination of the effects of encoding and the effects of mathematical misspecification of the test and omission of covariates. We further observe that there is a large proportion (almost 50%) of potentially alternative hypotheses. Given such a high number, there is potentially for numerous false positives if we were to reject the null using the $p$-value (or $z$-score) alone. Thus, we require FDR control in order to make more careful inference.

Using equations (11) and (12), we controlled the estimated $m$FDR at the $\beta = 0.1$ level by setting the threshold $c_{0.1} = 0.09986$. This resulted in 608685 of the voxels being declared significantly correlated with the seed, under FDR control, which equates to 21.60%. Figure 4 displays visualizations of the significant voxels at the perpendicular cross-sections intersecting the seed point $(x, y, z) = (125, 124, 64)$.

Upon inspection of Figure 4 we observe that significant correlation with the seed vector appears to be exhibited across the brain. The displays A2 and A3 in Figure 4 further show that the correlation appears to be symmetric between the two hemispheres. Furthermore, the correlation patterns appear in contiguous and smooth regions. The observations of whole-brain correlation with the bed nucleus of the stria terminalis are well supported in the literature. For example, similar connectivity observations were made by Dong et al. (2001) and Dong & Swanson (2006) in mouse studies, and by McMenamin & Pessoa (2015) and Torrisi et al. (2015) in human studies.

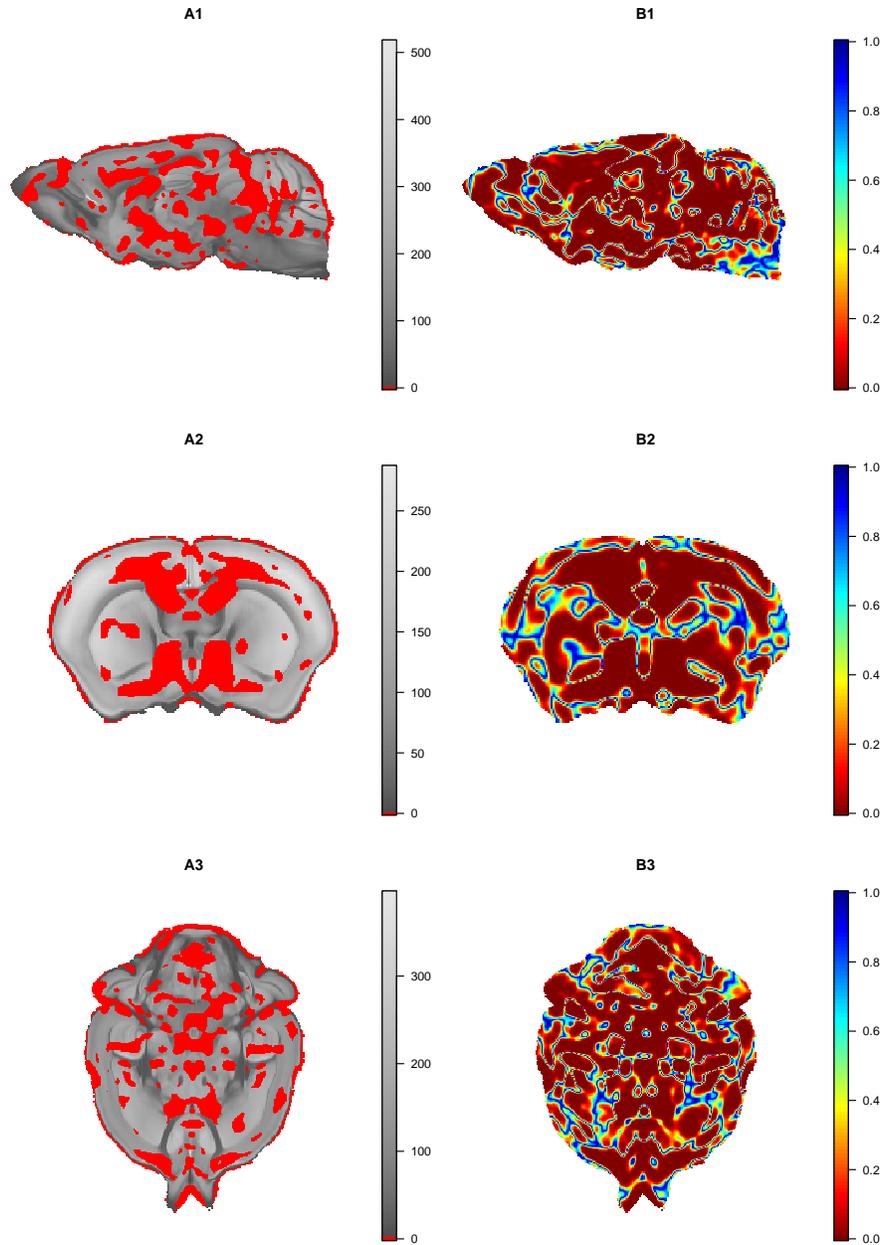

Figure 4: A1 and B1 display the anatomic background MRI intensities and $p$-values for the $x = 125$ slice, respectively. Similarly A2 and B2 display the respective quantities for the $y = 124$ slice, and A3 and B3 display the respective quantities for the $z = 64$ slice. In A1–A3, red voxels indicate those that are significant when controlled at the $\beta = 0.1$ FDR level.

# 6   Conclusions

We have presented an EB-based FDR control method for the mitigation of false positive results in multiple simultaneous hypothesis testing scenarios where only $p$-values or test statistics are available from the hypothesis tests, and when these $p$-values and test statistics are potentially precision-reduced via integer encodings. Due to the nature of the construction of our method, it is robust to situations where the hypothesis tests are misspecified or when there may be omitted covariates that have not been included in the testing procedure.

In order to handle the encoded testing data, we utilized a finite mixture model that can be estimated from binned data. We proved that the parameter vector of the mixture model can also be estimated consistently, even when the testing data may be correlated. A simulation study was used to demonstrate that our methodology was competitive with some popular methods in well-specified testing scenarios, and outperformed these methods when the testing data arise from misspecified tests.

Finally a brain imaging study of mice was conducted to demonstrate our methodology in practice. The study constituted a whole-brain voxel-based study of connectivity to the bed nucleus of the stria terminalis, consisting of $n = 2818191$ tests. The $p$-values for the study were obtained from a complex pipeline that resulted in a set of encoded values, which included zeros and ones. Furthermore, the $p$-values were correlated (due to the spatial nature of imaging and subsequent processing) and the hypothesis tests were conducted under mathematical assumptions that may have lead to misspecification. As such, the use of our methodology was most suitable for the study. As a result of the study, we found whole-brain correlation patterns that were consistent with those found in the literature.

Conducting FDR control under reduced precision computation is becoming more prevalent with the popularity of open-access neuroimaging repositories such as those that are cited in Eickhoff et al. (2016). Our methodology provides a simple and robust solution to performing inference when $p$-values are computed from test statistics that are stored in legacy data formats or are themselves originally compressed in a precision reduced manner.

## Acknowledgements

HDN is funded by Australian Research Council grant DE170101134. GJM is funded by a Discovery Project grant from the Australian Research Council Discovery.

## Appendix

### Proof of Proposition 1

In order to apply van der Vaart (1998, Thm. 5.14), We must check that (i) $\log \mathbb{P}(\boldsymbol{X}_1 = \boldsymbol{x}; \boldsymbol{\theta})$ is continuous for all values of $\boldsymbol{x}$, and that (ii) the uniform strong law of large numbers holds; that is

$$\sup_{\boldsymbol{\theta} \in \Theta} \left| \frac{1}{n} \sum_{i=1}^{n} \log \mathbb{P}(\boldsymbol{X}_i = \boldsymbol{x}; \boldsymbol{\theta}) - \mathbb{E} \log \mathbb{P}(\boldsymbol{X}_1 = \boldsymbol{x}; \boldsymbol{\theta}) \right| \overset{\text{a.s.}}{\to} 0.$$

Property (i) is simple to verify since $\mathbb{P}(\boldsymbol{X}_1 = \boldsymbol{x}; \boldsymbol{\theta})$ can be written as an integral of a smooth function for any $\boldsymbol{x}$. Thus it is continuous and its logarithm is also continuous. To establish property (ii), we utilize Andrews (1992, Thm. 4). This requires that $n^{-1} \sum_{i=1}^{n} \log \mathbb{P}(\boldsymbol{X}_1 = \boldsymbol{x}; \boldsymbol{\theta})$ converges to $\mathbb{E} \log \mathbb{P}(\boldsymbol{X}_1 = \boldsymbol{x}; \boldsymbol{\theta})$, pointwise, almost surely for any $\boldsymbol{\theta} \in \Theta$, and that $\mathbb{E} \sup_{\boldsymbol{\theta} \in \Theta} |\log \mathbb{P}(\boldsymbol{X}_1 = \boldsymbol{x}; \boldsymbol{\theta})| < \infty$. For any $\boldsymbol{\theta}$, the variance of $\log \mathbb{P}(\boldsymbol{X}_1 = \boldsymbol{x}; \boldsymbol{\theta})$ exists since it is a discrete random variable with only finite outcomes. Thus, we can apply the mixing continuous mapping theorem and the mixing strong law of large numbers (i.e. White, 2001, Thm 3.49 and Cor. 3.48), in order to obtain the pointwise convergence of $n^{-1} \sum_{i=1}^{n} \log \mathbb{P}(\boldsymbol{X}_1 = \boldsymbol{x}; \boldsymbol{\theta})$, almost surely. Next, we again note that $\log \mathbb{P}(\boldsymbol{X}_1 = \boldsymbol{x}; \boldsymbol{\theta})$ is a discrete random variable with finite outcomes for any finite $\boldsymbol{\theta}$. Therefore, the supremum and its expectation are also finite, since $\Theta$ contains only finite values. Therefore (ii) is verified and the proposition is proved.

*Remark* 3. We note that van der Vaart (1998, Thm. 5.14) only lists the requirement to check assumption (ii) for the proof above. However, the theorem also makes an implicit assumption that the data are independent. Under dependence, we require the additional assumption of the strong law of large numbers (i), as demanded by Andrews (1992, Thm. 4). Here, we utilize the TSE-1D form of the theorem (cf. Andrews, 1992, Eqn. 3.2).

Same, A. (2009). Grouped data clustering using a fast mixture-model-based algorithm. In *Proceedings of the 2009 IEEE International Conference on Systems, Man, and Cybernetics*.

Scott, D. W. (1979). On optimal and data-based histograms. *Biometrika*, 66, 605–610.

Seeley, W. W., Zhou, R. K. C. J., Miller, B. L., & Greicius, M. D. (2009). Neurodegenerative diseases target large-scale human brain networks. *Neuron*, 62, 42–52.

Sharda, M., Khundrakpam, B. S., Evans, A. C., & Singh, N. C. (2016). Disruption of structural covariance networks for language in autism is modulated by verbal ability. *Brain Structure and FUnction*, 221, 1017–1032.

Sled, J. G., Zijdenbos, A. P., & Evans, A. C. (1998). A nonparametric method for automatic correction of intensity nonuniformity in MRI data. *IEEE Transactions on Medical Imaging*, 17, 87–97.

Spokoiny, V. & Dickhaus, T. (2015). *Basics of Modern Mathematical Statistics*. New York: Springer.

Storey, J. D. (2002). A direct approach to false discovery rates. *Journal of the Royal Statistical Society Series B*, 64, 479–498.

Storey, J. D., Bass, A. J., Dabney, A., & Robinson, D. (2015). *qvalue: Q-value estimation for false discovery rate control*.

Storey, J. D. & Tibshirani, R. (2003). Statistical significance for genomewide studies. *Proceedings of the National Academy of Sciences*, 100, 9440–9445.

Sturges, H. A. (1926). The choice of a class interval. *Journal of the American Statistical Association*, 21, 65–66.
53

White, H. (1982). Maximum likelihood estimation of misspecified models. *Econometrica*, 50, 1–25.

White, H. (1994). *Estimation, Inference and Specification Analysis*. Cambridge: Cambridge University Press.

White, H. (2001). *Asymptotic Theory For Econometricians*. San Diego: Academic Press.

Wu, C. F. J. (1983). On the convergence properties of the EM algorithm. *Annals of Statistics*, 11, 95–103.

Wu, J. & Hamdan, H. (2013a). Bin-EM-CEM algorithms of general parsimonious Gaussian mixture models for binned data clustering. In *IEEE 9th International Conference on Computational Cybernetics*.

Wu, J. & Hamdan, H. (2013b). Model choice for binned-EM algorithms of fourteen parsimonious Gaussian mixture models by BIC and ICL criteria. In *IEEE International Conference on System Science and Engineering*.

Xie, J., Cai, T. T., Maris, J., & Li, H. (2011). Optimal false discovery rate control for dependent data. *Statistics and Its Interface*, 4, 417–430.

Yekutieli, D. (2008). False discovery rate control for non-positively regression dependent test statistics. *Journal of Statistical Planning and Inference*, 138, 405–415.

Yekutieli, D. & Benjamini, Y. (1999). Resampling-based false discovery rate controlling multiple test procedures for correlated test statistics. *Statistical Planning and Inference*, 82, 171–196.55